\pdfoutput=1
\documentclass[journal]{IEEEtran}      					   %

\IEEEoverridecommandlockouts  %
\usepackage[usenames,dvipsnames]{xcolor}
\usepackage{graphicx} %
\usepackage{times} %
\usepackage{amsmath,amssymb}
\usepackage{pstricks,pst-plot,psfrag}
\usepackage[per-mode=symbol]{siunitx}
\usepackage{datetime}
\usepackage{multirow}
\usepackage[capitalise,noabbrev]{cleveref}
\usepackage[]{mcode}
\usepackage{bbold}
\usepackage{balance}
\usepackage{stmaryrd}
\usepackage[acronym]{glossaries}
\usepackage{cite}
\usepackage{lettrine}
\usepackage{import}
\usepackage{tikz,tikz-cd}
\usepackage{mathtools}
\usepackage{dsfont}
\usepackage{booktabs}
\usepackage{bbm}
\usepackage{subcaption}
\usepackage{comment}
\usepackage{ifthen}
\usepackage{url}
\usepackage{nomencl}
\renewcommand{\nomgroup}[1]{
  \ifthenelse{\equal{#1}{C}}{\item[\textbf{Co-design-related symbols}]}{
  \ifthenelse{\equal{#1}{A}}{\item[\textbf{AMoD-related symbols}]}{
  \ifthenelse{\equal{#1}{D}}{\item[\textbf{\ldots}]}{
  \ifthenelse{\equal{#1}{O}}{\item[\textbf{Other symbols}]}{}}}}
}
\makenomenclature
\newboolean{proofs}

\newcommand{\cG}{\mathcal{G}}

\newcommand{\cM}{\mathcal{M}}

\newcommand{\cV}{\mathcal{V}}

\definecolor{lightblue}{rgb}{0.60784,0.76078,0.90196}
\definecolor{darkblue}{rgb}{0.26667,0.44706,0.76863}
\definecolor{lightgreen}{rgb}{0.66275,0.81569,0.55686}
\definecolor{darkgreen}{rgb}{0.43922,0.67843,0.27843}
\definecolor{orange}{rgb}{0.92941,0.49020,0.19216}
\definecolor{yellow}{rgb}{1.00000,0.75294,0.00000}
\definecolor{grey}{rgb}{0.64706,0.64706,0.64706}
\definecolor{purple}{rgb}{0.51373,0.23529,0.04706}
\newacronym{abk:amod}{AMoD}{Autonomous Mobility-on-Demand}
\newacronym{abk:iamod}{\mbox{I-AMoD}}{intermodal \gls{abk:amod}}
\newacronym{abk:av}{\mbox{AV}}{autonomous vehicle}

\newacronym{abk:bpr}{BPR}{Bureau of Public Roads}
\newacronym{abk:bev}{BEV}{Battery Electric Vehicle}
\newacronym{abk:ca}{CA}{congestion-aware}
\newacronym{abk:cara}{CARS}{congestion-aware routing scheme}
\newacronym{abk:cpo}{CPO}{complete partial order}
\newacronym{abk:cdp}{CDP}{co-design problem}
\newacronym{abk:cdpi}{CDPI}{co-design problem with implementation}
\newacronym{abk:dp}{DP}{design problem}
\newacronym{abk:dpi}{DPI}{design problem with implementation}
\newacronym{abk:mdpi}{MDPI}{monotone design problem with implementation}
\newacronym{abk:dcpo}{DCPO}{directed complete partial order}
\newacronym{abk:es}{ES}{e-scooter}
\newacronym{abk:ffcs}{FFCS}{free floating car sharing systems}
\newacronym{abk:fcm}{FCM}{fuel-cell moped}
\newacronym{abk:ghg}{GHG}{greenhouse gas}
\newacronym{abk:icev}{ICEV}{ Internal Combustion Engine Vehicle}
\newacronym{abk:kpi}{KPIs}{Key Performance Indicators}
\newacronym{abk:lw}{LW}{Lightweight}
\newacronym{abk:mm}{{$\mu$}M}{micromobility}
\newacronym{abk:mmveh}{{$\mu$}MV}{micromobility vehicle}
\newacronym{abk:mod}{MoD}{Mobility-on-Demand}
\newacronym{abk:mcdp}{MCDP}{Monotone Co-Design Problem}
\newacronym{abk:mcfp}{MCFP}{multi-commodity flow problem}
\newacronym{abk:moped}{M}{moped}
\newacronym{abk:nyc}{NYC}{New York City}
\newacronym{abk:poset}{poset}{partially ordered set}
\newacronym{abk:ptop}{P2P}{point-to-point}
\newacronym{abk:sb}{SB}{shared bike}
\newacronym{abk:spp}{SPP}{shortest path problem}
\newacronym{abk:kdspp}{k-dSPP}{k-disjoint \gls{abk:spp}}
\newacronym{abk:su}{SU}{Sport Utility}

\newcommand{\achievableSpeedVeh}{v_\mathrm{V}}\nomenclature[a100]{$\achievableSpeedVeh$}{Speed~\gls{abk:av}}
\newcommand{\achievableSpeedSco}{v_\mathrm{M}}\nomenclature[a101]{$\achievableSpeedSco$}{Speed~\gls{abk:mm}}
\newcommand{\antichain}{\mathsf{A}}\nomenclature[c2]{$\antichain \mathcal{P}$}{Set of antichains for poset $\mathcal{P}$}
\newcommand{\colorEdge}{c}\nomenclature[a1]{$\colorEdge$}{Edge-coloring maps}
\newcommand{\colorEdgeRoad}{c_\mathrm{R}}
\newcommand{\colorEdgeRoadVeh}{c_\mathrm{R,V}}
\newcommand{\colorEdgeRoadSco}{c_\mathrm{R,M}}
\newcommand{\colorEdgeSubway}{c_\mathrm{P}}
\newcommand{\colorEdgePedestrian}{c_\mathrm{W}}
\newcommand{\colorEdgeConnections}{c_\mathrm{C}}

\nomenclature[c1]{$\bot, \top$}{Bottom and top of a poset}
\nomenclature[c1]{$d$}{Design problem}

\nomenclature[a1]{$\rho$}{Revenue}
\nomenclature[c1]{$\mathcal{P}$}{Poset and powerset}
\nomenclature[a1]{$\mathcal{G}$}{Digraph}
\nomenclature[a1]{$\mathcal{V}$}{Set of vertices of a digraph}
\nomenclature[a1]{$\mathcal{A}$}{Set of arcs of a digraph}
\nomenclature[a1]{$q$}{Travel demand}

\newcommand{\colorMetrics}{\mathcal{S}}
\newcommand{\arcBaselineUsage}{u_{ij}}\nomenclature[a9]{$\arcBaselineUsage$}{Baseline usage for arc $\tup{i,j}$}

\newcommand{\arcEnergy}{e_{ij}}\nomenclature[a9]{$\arcEnergy$}{Energy consumption for arc $\tup{i,j}$}

\newcommand{\arcNormalCapacity}{k_{ij}}\nomenclature[a9]{$\arcNormalCapacity$}{Normal capacity of arc $\tup{i,j}$}

\newcommand{\arcSpeedLimit}{v_{{\mathrm{L},ij}}}\nomenclature[a9]{$\arcSpeedLimit$}{Speed limit for arc $\tup{i,j}$}

\newcommand{\arcLength}{s_{ij}}\nomenclature[a1]{$\arcLength$}{Length of arc $\tup{i,j}$}

\nomenclature[a1]{$s_\mathrm{cycle}$}{Length of cycle}
\newcommand{\arcSpeedVeh}{v_{\mathrm{V},ij}}\nomenclature[a9]{$\arcSpeedVeh$}{Speed of \glspl{abk:av} on arc $\tup{i,j}$}

\newcommand{\arcSpeedWalking}{v_{\mathrm{W}}}\nomenclature[a8]{$\arcSpeedWalking$}{Walking speed}

\newcommand{\arcSpeedSco}{v_{\mathrm{M},ij}}
\nomenclature[a9]{$\arcSpeedSco$}{Speed of \glspl{abk:mmveh} on arc $\tup{i,j}$}
\newcommand{\arcTime}{t_{ij}}\nomenclature[a2]{$\arcTime$}{Travel time of arc $\tup{i,j}$}

\newcommand{\averageTravelTime}{t_\mathrm{avg}}\nomenclature[a3]{$\averageTravelTime$}{Average travel time}

\newcommand{\costFixVeh}{C_\mathrm{V,f}}\nomenclature[a15]{$\costFixVeh$}{Vehicle fix cost}

\newcommand{\costFixSco}{C_\mathrm{M,f}}\nomenclature[a16]{$\costFixSco$}{\gls{abk:mmveh} fix cost}

\newcommand{\costFixTrain}{C_\mathrm{S,f}}\nomenclature[a16]{$\costFixTrain$}{Subway fix cost}

\newcommand{\costOpTrain}{C_\mathrm{S,o}}\nomenclature[a16]{$\costOpTrain$}{Subway operational cost}

\newcommand{\costOpVeh}{C_\mathrm{V,o}}\nomenclature[a16]{$\costOpVeh$}{\gls{abk:av} operational cost}

\newcommand{\costOpSco}{C_\mathrm{M,o}}\nomenclature[a16]{$\costOpSco$}{\gls{abk:mmveh} operational cost}

\newcommand{\costTot}{C_\mathrm{tot}}\nomenclature[a16]{$\costTot$}{Total cost}
\newcommand{\costVeh}{C_\mathrm{V}}
\newcommand{\costSub}{C_\mathrm{S}}
\newcommand{\costSco}{C_\mathrm{M}}
\newcommand{\distanceVeh}{s_\mathrm{V,tot}}\nomenclature[a16]{$\distanceVeh$}{Total distance \gls{abk:av}}
\newcommand{\distanceSco}{s_\mathrm{M,tot}}\nomenclature[a16]{$\distanceSco$}{Total distance \gls{abk:mmveh}}

\newcommand{\emissionsVeh}{m_\mathrm{CO_2,V,tot}}\nomenclature[a15]{$\emissionsVeh$}{Emissions \gls{abk:av}}
\newcommand{\emissionsSco}{m_\mathrm{CO_2,M,tot}}\nomenclature[a15]{$\emissionsSco$}{Emissions \gls{abk:mmveh}}
\newcommand{\emissionsTot}{m_\mathrm{CO_2,tot}}
\newcommand{\emissionsSubTrain}{m_\mathrm{CO_2,S}}
\newcommand{\emissionsSub}{m_\mathrm{CO_2,S,tot}}\nomenclature[a15]{$\emissionsSub$}{Emissions subway}
\newcommand{\flow}[2]{f_m\left(#1,#2\right)}\nomenclature[a4]{$\flow{i}{j}$}{Flow of customers per unit time on arc $\tup{i,j}$}
\newcommand{\flowRebaVeh}[2]{f_{0,\mathrm{V}}\left(#1,#2\right)}\nomenclature[a5]{$\flowRebaVeh{i}{j}$}{Flow of empty \glspl{abk:av} on arc $\tup{i,j}$}

\newcommand{\flowRebaSco}[2]{f_{0,\mathrm{M}}\left(#1,#2\right)}\nomenclature[a6]{$\flowRebaSco{i}{j}$}{Flow of empty \glspl{abk:mmveh} on arc $\tup{i,j}$}

\newcommand{\flowTotVeh}[2]{f_{\mathrm{tot},\mathrm{V}}\left(#1,#2\right)}\nomenclature[a7]{$\flowTotVeh{i}{j}$}{Total flow of \glspl{abk:av} on arc $\tup{i,j}$}

\newcommand{\flowTotSco}[2]{f_{\mathrm{tot},\mathrm{M}}\left(#1,#2\right)}\nomenclature[a7]{$\flowTotSco{i}{j}$}{Total flow of \glspl{abk:mmveh} on arc $\tup{i,j}$}

\newcommand{\freqTrain}{\varphi_j}\nomenclature[a15]{$\freqTrain$}{Subway service frequency}
\newcommand{\freqTrainBaseline}{\varphi_{j,\mathrm{base}}}\nomenclature[a15]{$\freqTrainBaseline$}{Subway baseline service frequency}
\newcommand{\lifeVeh}{l_\mathrm{V}}\nomenclature[a15]{$\lifeVeh$}{Lifetime \gls{abk:av}}
\newcommand{\lifeTrain}{l_\mathrm{S}}\nomenclature[a15]{$\lifeTrain$}{Lifetime subway train}
\newcommand{\lifeSco}{l_\mathrm{M}}\nomenclature[a15]{$\lifeSco$}{Lifetime \gls{abk:mmveh}}
\newcommand{\numberFleetVeh}{n_\mathrm{V,max}}\nomenclature[a15]{$\numberFleetVeh$}{Fleet size \glspl{abk:av}}
\newcommand{\numberFleetSco}{n_\mathrm{M,max}}\nomenclature[a15]{$\numberFleetSco$}{Fleet size \glspl{abk:mmveh}}

\newcommand{\numberFleetUsedVeh}{n_\mathrm{V,u}}\nomenclature[a15]{$\numberFleetUsedVeh$}{Used \glspl{abk:av}}

\newcommand{\numberFleetUsedSco}{n_\mathrm{M,u}}
\nomenclature[a15]{$\numberFleetUsedSco$}{Used \glspl{abk:mmveh}}
\newcommand{\numberFleetTrain}{n_\mathrm{S}}\nomenclature[a15]{$\numberFleetTrain$}{Fleet size subway}
\newcommand{\numberFleetTrainBaseline}{n_\mathrm{S,base}}
\newcommand{\power}[1]{\mathcal{P}(#1)}
\newcommand{\prov}{\mathsf{prov}}\nomenclature[c2]{$\prov$}{Implementation-to-functionality map}
\newcommand{\rplusbar}{\overline{\mathbb{R}}_{\geq 0}}

\newcommand{\req}{\mathsf{reqs}}\nomenclature[c2]{$\req$}{Implementation-to-resourses map}

\newcommand{\traveltime}{t}
\newcommand{\timePedestrianSubway}{\traveltime_{\mathrm{WS}}}
\newcommand{\timeSubwayPedestrian}{\traveltime_{\mathrm{SW}}}
\newcommand{\timePedestrianRoadVeh}{\traveltime_{\mathrm{WV}}}
\newcommand{\timeRoadVehPedestrian}{\traveltime_{\mathrm{VW}}}
\newcommand{\timePedestrianRoadSco}{\traveltime_{\mathrm{WM}}}
\newcommand{\timeRoadScoPedestrian}{\traveltime_{\mathrm{MW}}}
\newcommand{\setOfArcs}{\mathcal{A}}

\newcommand{\setOfArcsRoad}{\mathcal{A}_{\mathrm{R}}}
\newcommand{\setOfArcsSubway}{\mathcal{A}_{\mathrm{P}}}
\newcommand{\setOfArcsPedestrian}{\mathcal{A}_{\mathrm{W}}}
\newcommand{\setOfArcsCommute}{\mathcal{A}_{\mathrm{C}}}
\newcommand{\setOfArcsRoadVeh}{\mathcal{A}_{\mathrm{R,V}}}
\newcommand{\setOfArcsRoadSco}{\mathcal{A}_{\mathrm{R,M}}}
\newcommand{\setOfFunctionalities}[1]{\F{\mathcal{F}_{#1}}}
\nomenclature[c1]{$\setOfFunctionalities{}$}{Functionalities}
\nomenclature[c1]{$\setOfResources{}$}{Resources}
\newcommand{\setOfFunctionalitiesOp}[1]{\F{\mathcal{F}_{#1}}^{\mathrm{op}}}
\definecolor{dpgreen}{rgb}{0.0, 0.5, 0.0}
\newcommand{\F}[1]{\textcolor{dpgreen}{#1}}

\newcommand{\GraphRoad}{\mathcal{G}_\mathrm{R}}
\newcommand{\GraphSubway}{\mathcal{G}_\mathrm{P}}
\newcommand{\GraphPedestrian}{\mathcal{G}_\mathrm{W}}
\newcommand{\GraphRoadVeh}{\mathcal{G}_\mathrm{R,V}}
\newcommand{\GraphRoadSco}{\mathcal{G}_\mathrm{R,M}}
\newcommand{\setOfImplementations}[1]{\mathcal{I}_{#1}}
\newcommand{\posA}{P}
\newcommand{\posB}{Q}
\newcommand{\tup}[1]{\langle #1 \rangle}
\definecolor{dpred}{rgb}{0.7, 0.0, 0.0}
\newcommand{\R}[1]{\textcolor{dpred}{#1}}
\newcommand{\setOfResources}[1]{\R{\mathcal{R}_{#1}}}
\newcommand{\setofGraphs}{\mathbf{G}}\nomenclature[a15]{$\setofGraphs$}{Set of graphs}

\newcommand{\posreals}{\mathbb{R}_{\geq 0}}

\ifPDFTeX
\newcommand{\tickar}{\begin{tikzcd}[baseline=-0.5ex,cramped,sep=small,ampersand replacement=\&]{}\ar[r,tick]\&{}\end{tikzcd}}
\else
\newcommand{\tickar}{\nrightarrow}
\fi
\newcommand{\setOfVertices}{\mathcal{V}}

\newcommand{\setOfVerticesRoad}{\mathcal{V}_{\mathrm{R}}}
\newcommand{\setOfVerticesRoadVeh}{\mathcal{V}_{\mathrm{R,V}}}
\newcommand{\setOfVerticesRoadSco}{\mathcal{V}_{\mathrm{R,M}}}
\newcommand{\setOfVerticesSubway}{\mathcal{V}_{\mathrm{P}}}
\newcommand{\setOfVerticesPedestrian}{\mathcal{V}_{\mathrm{W}}}
\newcommand{\multigraph}{\mathcal{G}}

\newcommand{\op}{^{\mathrm{op}}}
\newcommand{\arc}{\tup{i,j}}

\newcommand{\bool}[1]{\mathds{1}_{#1}}

\newcommand{\true}[1]{\mathtt{T}}

\nomenclature[a15]{$\mathrm{red}$}{Graph reduction maps}

\nomenclature[a15]{$\pi$}{Projection maps}

\usepackage{amsthm}

\newtheorem{theorem}{Theorem}[section]
\newtheorem{lemma}[theorem]{Lemma}
\theoremstyle{definition}

\newtheorem{definition}[theorem]{Definition}
\newtheorem{example}[theorem]{Example}
\newtheorem{problem}{Problem}

\theoremstyle{remark}
\newtheorem*{remark}{Remark}

\newcommand{\changed}[1]{{#1}}

\newcommand{\bischanged}[1]{\textcolor{black}{#1}}

\makeatletter
    \if@cref@capitalise
        \crefname{subsection}{Section}{Sections}
        \crefname{subsubsection}{Section}{Sections}
        \crefname{assump}{Assumption}{Assumptions}
        \crefname{problem}{Problem}{Problems}
    \else
        \crefname{subsection}{section}{sections}
        \crefname{subsubsection}{section}{sections}
        \crefname{assump}{assumption}{assumptions}
        \crefname{problem}{problem}{problems}
    \fi
\makeatother

\DeclareSIUnit{\eur}{\euro}
\DeclareSIUnit{\usd}{USD}
\DeclareSIUnit{\mph}{mph}
\DeclareSIUnit{\month}{month}
\DeclareSIUnit{\year}{year}
\DeclareSIUnit{\million}{Mil}
\DeclareSIUnit{\mile}{mile}
\DeclareSIUnit{\car}{car}
\DeclareSIUnit{\train}{train}
\DeclareSIUnit{\mmveh}{\text{$\mu$}MV}
\DeclareSIUnit{\nounit}{-}
\usetikzlibrary{positioning,intersections, hobby, patterns, calc, decorations.pathmorphing, decorations.markings, shadows,shapes, cd,arrows.meta, fit,quotes}

\tikzset{
   tick/.style={postaction={
      decorate,
      decoration={markings, mark=at position 0.5 with {\draw[-] (0,.4ex) -- (0,-.4ex);}}}
   }
}

\tikzstyle{block} = [draw, rectangle, minimum height=2em, minimum width=3em,font=\bfseries,rounded corners,thick]
\tikzstyle{block1} = [draw, rectangle, minimum height=1.5em, minimum width=2.5em]
\tikzstyle{blockDyn} = [draw, rectangle, minimum height=2.5em, minimum width=3.5em, align=center, inner sep=10pt, thick, fill=white, copy shadow={draw=black,fill=black,opacity=1,shadow xshift=0.5ex,shadow yshift=-0.5ex}]
\tikzstyle{blockAlg} = [draw, rectangle, minimum height=1.5em, minimum width=2.5em, align=center, inner sep=10pt, thick]
\tikzstyle{sum} = [draw,circle]
\tikzstyle{nodePre} = [circle, draw,inner sep=1pt,node contents={$\preceq$},thick]
\tikzstyle{nodePreEmpty} = [circle, draw,inner sep=1pt,thick]
\tikzstyle{nodePos} = [circle, draw,inner sep=1pt,node contents={$\posceq$},thick]
\tikzstyle{nodeProd} = [rectangle, draw,inner sep=4pt,node contents={$\times$},rounded corners,thick]
\tikzstyle{nodeSum} = [rectangle, draw,inner sep=4pt,node contents={$\mathbf{+}$},rounded corners,thick]

\definecolor{DPgreen}{RGB}{34,139,34}

\usetikzlibrary{positioning,intersections, hobby, patterns, calc, decorations.pathmorphing, decorations.markings, shadows,shapes, cd,arrows.meta, fit,quotes}

\tikzset{
   tick/.style={postaction={
      decorate,
      decoration={markings, mark=at position 0.5 with {\draw[-] (0,.4ex) -- (0,-.4ex);}}}
   }
}
\tikzstyle{block} = [draw, rectangle, minimum height=2em, minimum width=3em,font=\bfseries,rounded corners,thick]
\tikzstyle{block} = [draw, rectangle, minimum height=2em, minimum width=3em]
\tikzstyle{block1} = [draw, rectangle, minimum height=1.5em, minimum width=2.5em]
\tikzstyle{blockDyn} = [draw, rectangle, minimum height=2.5em, minimum width=3.5em, align=center, inner sep=10pt, thick, fill=white, copy shadow={draw=black,fill=black,opacity=1,shadow xshift=0.5ex,shadow yshift=-0.5ex}]
\tikzstyle{blockAlg} = [draw, rectangle, minimum height=1.5em, minimum width=2.5em, align=center, inner sep=10pt, thick]
\tikzstyle{sum} = [draw,circle]

\tikzstyle{nodePre} = [circle, draw,inner sep=1pt,node contents={$\preceq$},thick]
\tikzstyle{nodePreEmpty} = [circle, draw,inner sep=1pt,thick]
\tikzstyle{nodePos} = [circle, draw,inner sep=1pt,node contents={$\posceq$},thick]
\tikzstyle{nodeProd} = [rectangle, draw,inner sep=4pt,node contents={$\times$},rounded corners,thick]
\tikzstyle{nodeSum} = [rectangle, draw,inner sep=4pt,node contents={$\mathbf{+}$},rounded corners,thick]

\definecolor{red}{rgb}{0.75, 0.0, 0.0}

\tikzset{fcname/.store in =\fcname, fcname={}}
\tikzset{funame/.store in =\funame, funame={}}
\tikzset{rcname/.store in =\rcname, rcname={}}
\tikzset{runame/.store in =\runame, runame={}}
\tikzset{whereres/.store in =\whereres, whereres=0.5}
\tikzset{wherefun/.store in =\wherefun, wherefun=0.5}
\tikzset{relres/.store in =\relres, relres={above}}
\tikzset{relfun/.store in =\relfun, relfun={above}}
\tikzset{posres/.store in =\posres, posres=1}
\tikzset{posfun/.store in =\posfun, posfun=1}
\tikzset{loos/.store in =\loos, loos=2}
\tikzset{feedback/.store in =\feedback, feedback=0}
\tikzset{
   DP/.style={%
      label/.style={
         font=\everymath\expandafter{\the\everymath\scriptstyle},
         inner sep=5pt,
         node distance=2pt and -2pt},
      semithick,
      node distance=1 and 1,
      rconn/.style={color=white,opacity=0.0,postaction={decorate}, shorten <=3.2pt, shorten >= 0.8,
      decoration={markings, 
      mark= at position 0 with {
               \coordinate (a);
      },
      mark=at position .5 with
      {
              \ifthenelse{\equal{\feedback}{1}}{\def\angleOut{90}\def\angleIn{90}}{\def\angleOut{0}\def\angleIn{180}}    
              \coordinate (b);
              \draw[dashed,dpred,opacity=1.0] (a) to[out=\angleOut,in=\angleIn,looseness=\loos] 
              node[pos=\posres,\relres=\whereres mm,dpred,opacity=1,fill=white,inner sep=1pt,outer sep=1pt]{\footnotesize{\rcname}} (b);
      },
      mark= at position 1 with 
      {
             \ifthenelse{\equal{\feedback}{1}}{\def\angleOut{0}\def\angleIn{0}}{\def\angleOut{180}\def\angleIn{0}} 
              \ifthenelse{\equal{\feedback}{1}}{\def\symbol{\succeq}}{\def\symbol{\preceq}} 
              \coordinate (c);
              \draw[dpgreen,opacity=1.0] (c) to[out=\angleOut,in=\angleIn,looseness=\loos]
              node[pos=\posfun,\relfun=\wherefun mm,dpgreen,opacity=1,fill=white,inner sep=1pt,outer sep=1pt]{\footnotesize{\fcname}} (b){}; %
              \node[draw,circle,inner sep=0.5pt,color=black,fill=white,opacity=1.0] at (b) (nodepreceq) {$\symbol$}; 
      }
      }},
      runconn/.style={color=dpred,dashed,postaction={decorate},
      decoration={markings,
      mark= at position 1 with {
              \coordinate (a);
              \draw[dpred,opacity=1.0,dashed] ($(a)+(0.05,0)$) --++ (0.5,0) node[\relres,pos=\posres]{\footnotesize{\runame}};}
      }
      },
      funconn/.style={color=white,postaction={decorate},
      decoration={markings,
      mark= at position 0 with {
      \coordinate (a);
      \draw[dpgreen] ($(a)+(-0.05,0)$) -- ($(a)+(-0.5,0)$) node[\relfun, pos=\posfun]{\footnotesize{\funame}};}
      }
      },
      execute at begin picture={\tikzset{
         x=\dpx, y=\dpy,
         every fit/.style={inner xsep=\dpx, inner ysep=\dpy}}}
      },
   dpx/.store in=\dpx,
   dpx = 1.5cm,
   dpy/.store in=\dpy,
   dpy = 1.5ex,
   dp port sep/.store in=\dpportsep,
   dp port sep=2,
   dp port length/.store in=\dpportlen,
   dp port length=4pt,
   dp min width/.store in=\dpminwidth,
   dp min width=0.5cm,
   dp rounded corners/.store in=\dpcorners,
   dp rounded corners=2pt,
   dp small/.style={dp port sep=1, dp port length=2.5pt, dpx=.4cm, dp min width=.4cm, dpy=.7ex},
   dp/.code 2 args={%
      \pgfmathsetlengthmacro{\dpheight}{\dpportsep * (max(#1,#2)) * \dpy}
      \pgfkeysalso{draw,%
        minimum width=\dpminwidth,%
        minimum height=\dpheight,%
        font=\bfseries,
        outer sep=0pt,%
        inner sep=5pt,%
        rounded corners=\dpcorners,
        thick,
        prefix after command={\pgfextra{\let\fixname\tikzlastnode}},
        append after command={\pgfextra{\draw
            \ifnum #1=0{} \else foreach \i in {1,...,#1} { 
            ($(\fixname.north west)!{\i/(#1+1)}!(\fixname.south west)$) +(0,0) node[solid,left,circle,color=dpgreen,draw,fill=dpgreen,scale=0.3] {} coordinate (\fixname_fun\i) -- +(0,0) coordinate (\fixname_fun\i')}\fi %
            \ifnum #2=0{} \else foreach \i in {1,...,#2} {
            ($(\fixname.north east)!{\i/(#2+1)}!(\fixname.south east)$) +(0,0) coordinate (\fixname_res\i') -- +(0,0) node[solid,right,circle,color=dpred,draw,fill=dpred,scale=0.3] {} coordinate (\fixname_res\i)}\fi;
         }}}
         },
      dp name/.style={append after command={\pgfextra{\node[label=center,inner sep=2pt,fill=white] at (\fixname) {\textbf{#1}};}}}
   }

\title{
Co-Design to Enable User-Friendly Tools to Assess the Impact of Future Mobility Solutions
}

\author{Gioele Zardini$^{1}$, Nicolas Lanzetti$^{2}$, Andrea Censi$^1$, Emilio Frazzoli$^{1}$, and Marco Pavone$^3$ 
\thanks{$^1$Institute for Dynamic Systems and Control,
        ETH Z\"urich, Z\"urich (ZH), Switzerland
        {\tt \{gzardini,acensi,emilio.frazzoli\}@ethz.ch}}%
\thanks{$^2$Automatic Control Laboratory,
        ETH Z\"urich, Z\"urich (ZH), Switzerland,
        {\tt lnicolas@ethz.ch}}%
\thanks{$^3$Department of Aeronautics and Astronautics, Stanford University, Stanford (CA), United States,
        {\tt pavone@stanford.edu}}
\thanks{This work was supported by the National Science Foundation under CAREER Award CMMI-1454737, and by the Swiss National Science Foundation under NCCR Automation, grant agreement 51NF40\_180545. This article solely reflects the opinions and conclusions of its authors and not NSF, NCCR, or any other entity.}
}

\newif\ifextendedversion %
\extendedversiontrue

\makeatletter
\newcommand\Min{\@tempcnta=\mathcode`\m\relax
\mathcode`\m=\mathcode`\M\min\mathcode`\m=\@tempcnta\relax}
\makeatother

\begin{document}

\setboolean{proofs}{false}
\maketitle
\begin{abstract}
The design of future mobility solutions and the design of the mobility systems they enable are closely coupled. 
Indeed, knowledge about the intended service of novel mobility solutions would impact their design and deployment process, whilst insights about their technological development could significantly affect transportation management policies.
This requires tools to study such a coupling and co-design mobility systems in terms of different objectives.
We present a framework to address such co-design problems, leveraging a mathematical theory of co-design to frame and solve the problem of designing and deploying an intermodal mobility system, whereby autonomous vehicles service travel demands jointly with micromobility solutions and public transit, in terms of fleets sizing, vehicle characteristics, and public transit service frequency. 
Our framework is modular and compositional, allowing one to describe the design as the interconnection of simple components and to tackle it from a systemic perspective.
Moreover, it requires general monotonicity assumptions and naturally handles multiple objectives, delivering rational, actionable solutions for policy makers.
\bischanged{We showcase our methodology in a case study of Washington D.C., USA.}
Our work suggests the possibility to create user-friendly optimization tools to systematically assess costs and benefits of interventions, and to inform policy-making in the future.
\end{abstract}
\section{Introduction}\label{sec:introduction}
Current transportation systems are undergoing dramatic mutations, arising from the deployment of novel mobility solutions, such as \glspl{abk:av} and \gls{abk:mm} systems.
\changed{New mobility paradigms promise to drastically reduce negative externalities produced by the transportation system, such as emissions, travel time, parking spaces and, critically, fatalities (for a review on the subject, refer to~\cite{zardinilanzettiAR2021}).}
\changed{However, industrial experience shows that the current design process for new mobility solutions often suffers from the lack of clear, specific requirements in terms of the service they will be providing~\cite{Yigitcanlar2019}.} 
Yet, knowledge about their intended service (e.g., last-mile versus point-to-point travel) might dramatically impact how vehicles are designed and significantly ease their development process. 
For instance, if for a given city we knew that for an effective on-demand mobility system \glspl{abk:av} only need to drive up to \SI{30}{\mph} and only on relatively \changed{simple} roads, their design would be greatly simplified and their deployment could be accelerated. 
Furthermore, from the system-level perspective of transportation management, knowledge about the trajectory of technology development for new mobility solutions would certainly impact decisions on future infrastructure investments and provisions of service. 
In other words, the design of future mobility solutions and the design of a mobility system leveraging them are intimately \emph{coupled}. 
This calls for methods to reason about such a coupling, and in particular to \emph{co-design} the invidual mobility solutions and the associated mobility systems. 
\changed{A key requirement in this context is to be able to account for a range of heterogeneous objectives that are often not directly comparable (consider, for instance, travel time, public expense, and externalities), to formulate hierarchical design problems involving different disciplines, and to solve them in a computationally tractable manner.}

\changed{Accordingly, the goal of this paper is to lay the foundations for a framework through which one can systematically co-design future mobility systems.}
Specifically, we show how to leverage a recently developed monotone theory of co-design~\cite{Censi2015,Censi2017b, Censi2017,censi2022}, which provides a general methodology to co-design complex systems in a modular and compositional fashion~\cite{zardiniIROS2020,zardiniECC21}.
This tool delivers the set of rational design solutions lying on the Pareto front, allowing one to reason about costs and benefits of the individual design options.
The framework is instantiated in the setting of co-designing intermodal mobility systems~\cite{SalazarLanzettiEtAl2019}, whereby fleets of self-driving vehicles provide on-demand mobility jointly with fleets of \glspl{abk:mmveh} such as \glspl{abk:es}, \glspl{abk:sb}, \glspl{abk:moped} and \glspl{abk:fcm}, and public transit.
Aspects that are subject to co-design include fleet sizes, vehicle-specific characteristics for \glspl{abk:av} and \glspl{abk:mmveh}, and service features, such as public transit service frequency, prices, and serviced networks.

\subsection{Related Literature}
Our work lies at the interface of the design of public transportation services and the design of novel mobility solutions.

The first research stream is reviewed in~\cite{FarahaniMiandoabchiEtAl2013,GuihaireHao2008, LODER2022113} and comprises \emph{strategic} long-term infrastructure modifications and \emph{operational} short-term scheduling.
The joint design of traffic network topology and control infrastructure has been presented in~\cite{CongDeSchutterEtAl2015, LUO2021100045}.
Public transportation scheduling has been solved jointly with the design of the transit network by optimizing passengers' and operators' costs in~\cite{Arbex2015}, the satisfied demand in~\cite{Sun2014}, and the energy consumptions of the system in~\cite{Su2013}.
However, these works mainly focus on a single infrastructure (road network or public transportation), and do not consider its joint design with new mobility solutions.

The research on novel mobility solutions mainly pertains to \glspl{abk:av}, \gls{abk:amod} systems, and \gls{abk:mm}.
The research on design of \gls{abk:amod} systems is thoroughly reviewed in~\cite{zardinilanzettiAR2021} and references therein, and mainly concerns their fleet sizing.
In this regard, existing studies range from simulation-based approaches~\cite{BarriosGodier2014,FagnantKockelman2018,Vazifeh2018,Boesch2016,MeghjaniEtAl2018,narayan2021fleet} to analytical methods~\cite{SpieserTreleavenEtAl2014}.
In~\cite{ZhangSheppardEtAl2018}, the fleet size and the charging infrastructure of an \gls{abk:amod} system are jointly designed, and the arising design problem is formulated as a mixed integer linear program.
In~\cite{BeaujonTurnquist1991}, the fleet sizing problem is solved together with the vehicle allocation problem.
\changed{Furthermore,~\cite{Wallar2019} proposes a framework to jointly design the \gls{abk:amod} fleet size and its composition.} 
\changed{More recently, the joint design of multi-modal transit networks and \gls{abk:amod} systems was formulated in~\cite{PintoHylandEtAl2019} as a bilevel optimization problem and solved with heuristics, and coupled with infrastructure design in~\cite{seo2021multi}, using multi-objective linear optimization.}
Overall, the problem-specific structure of existing design methods for \gls{abk:amod} systems is \changed{often} not amenable to a modular and compositional problem formulation. 
Furthermore, key \gls{abk:av} characteristics, such as the achievable speed, are not considered.
Research on the design and impact of \gls{abk:mm} solutions has been reviewed in~\cite{ShaheenEtAl2019}, which focuses on the urban deployment of \glspl{abk:sb} and \glspl{abk:es}. In particular,~\cite{GhamamiEtAl2018} presents a design framework for a multi-modal public transportation system, including various \gls{abk:mm} solutions and buses, optimizing user preferences and social costs.
Fleet deployment models are analyzed in~\cite{Lu2016,YanEtAl2018, luo2021optimal, chow2014symbiotic}. 
The optimal allocation of \glspl{abk:sb} in a city is studied through mathematical programming models in~\cite{Lu2016}, and solved through stochastic optimization in~\cite{YanEtAl2018,luo2021optimal}. Finally,~\cite{KondorEtAl2019} explores the impact of \gls{abk:mm} on urban planning and identifies strategies to increase \glspl{abk:mmveh} utilization.

\changed{At a higher abstraction level, the problem we are trying to solve matches some of the principles of collaborative engineering~\cite{lu2007scientific, zhang2008exploring}, which, while providing interesting insights, do not offer a mathematical theory and a scalable computational framework to deal with hierarchical, multi-objective design problems}.

In conclusion, to the best of the authors' knowledge, the existing design frameworks for mobility systems have a fixed problem-specific structure, and therefore do not permit to \emph{co-design} the mobility infrastructure in a modular and compositional manner. 
\changed{Moreover, previous works neither capture important aspects of future mobility systems, such as the interactions among different transportation modes, nor specific design parameters of novel mobility solutions, as for instance the level of autonomy and the serviced network of \glspl{abk:av}, in a computationally tractable and compositional way.}

\subsection{Statement of contributions}
In this paper, we lay the foundations for the systematic study of the design of future mobility systems. 
Specifically, we leverage a mathematical theory of co-design~\cite{Censi2015, censi2022} to devise a framework to study the design of intermodal mobility systems in terms of mobility solutions and fleet characteristics, enabling the computation of the \emph{rational} solutions lying on the Pareto front of minimal travel time, transportation costs, and externalities.
Our framework paves the way to structure the design problem in a modular way, in which each different transportation option can be ``plugged in'' in a larger model. 
\changed{Each model has minimal assumptions: Rather than properties such as linearity, continuity, or convexity, we ask for very general monotonicity assumptions, proven to be reasonable in the present paper and in previous works~\cite{Censi2017,zardiniIROS2020,zardiniECC21}.}
\changed{For example, we assume that the cost of automation of an \gls{abk:av} does not decrease with the increase of the speed achievable by the \gls{abk:av}.}
We are able to obtain the full Pareto front of \emph{rational} solutions or, given policies, to weigh incomparable costs (such as travel time and emissions) and to present actionable information to the stakeholders of the mobility ecosystem.
\changed{We consider the case study of Washington~D.C., USA, to showcase our methodology.}
We illustrate how, given the model, we can easily formulate and answer several questions regarding the introduction of new technologies and investigate possible infrastructure interventions.
This article significantly extends the preliminary material previously presented in~\cite{ZardiniLanzettiEtAl2020b,ZardiniLanzettiEtAl2020}. 
In particular, we first broaden and extend the formal presentation of the mathematical theory of co-design, and detail its application in this work by formalizing the introduced design problems (including proofs, new insights, and generalizability of the approach).
Second, we extend the discussion of the literature, including recent research pertaining to the co-design of future mobility systems, with a focus on \gls{abk:mm}.
\bischanged{Third, we showcase the modularity of our framework by including the design of \gls{abk:mm} solutions (both at the vehicle and at the fleet level) in the future mobility co-design problem, and evaluate their impact on the transportation system for the case study of Washington D.C., USA, whereby we leverage the network of the city, as well as demand datasets.}
Furthermore, we include a study of pricing strategies in the mobility co-design problem, highlighting the flexibility of the proposed framework.
Finally, we extend our case studies with further scenarios and provide new managerial insights.

\subsection{Organization of the paper}
The remainder of this paper is structured as follows: \cref{sec:co-designtheory} reviews the mathematical background on which our framework is based. 
\cref{sec:codesignav} presents models for future mobility systems and related co-design problems, by introducing the single \glspl{abk:dp} and their interconnection forming a \gls{abk:cdp}.
\bischanged{We showcase our methodology with several case studies for Washington D.C., USA, in \cref{sec:results}.}
\cref{sec:conclusion} concludes the paper with a discussion and an overview on future research directions.
\bischanged{Nomenclature is available in the appendix.}
\changed{%
\section{Monotone Co-Design Theory}
\label{sec:co-designtheory}

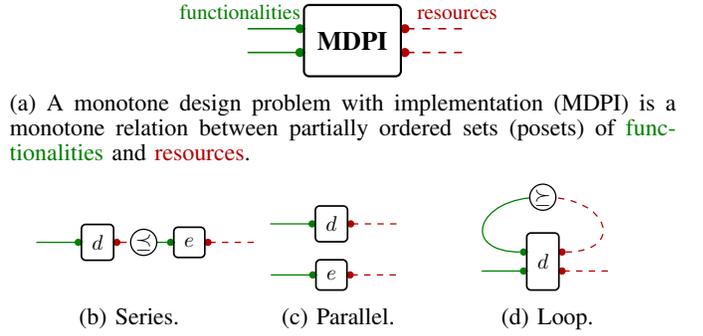
\begin{figure}[t]
\begin{subfigure}{\columnwidth}
    \begin{center}
    \begin{tikzpicture}[DP]
            \node[dp={2}{2}] (cnt) {MDPI};
            \draw[runconn, runame={resources}, relres=above,posres=0.9] (cnt_res1){};
            \draw[runconn, runame={}, relres=above,posres=0.9] (cnt_res2){};
            \draw[funconn, funame={functionalities},relfun=above,posfun=1.15] (cnt_fun1){};
            \draw[funconn, funame={},relfun=above,posfun=1.15] (cnt_fun2){};
\end{tikzpicture}
    \subcaption{A \gls{abk:mdpi} is a monotone relation between \glspl{abk:poset} of \F{functionalities} and \R{resources}. \label{fig:mathcodesign}}
    \end{center}
    \end{subfigure}
\begin{center}
\begin{subfigure}[b]{0.3\columnwidth}
\centering
\scalebox{0.8}{\begin{tikzpicture}[DP]
    \node[dp={1}{1}] (f) {$d$};
    \node[dp={1}{1}, right=1cm of f] (g) {$e$};
    \draw[rconn, rcname={}, fcname={}] (f_res1)  to (g_fun1);
    \draw[runconn, runame={}] (g_res1);
    \draw[funconn, funame={}] (f_fun1);
\end{tikzpicture}}
\subcaption{Series.}
\end{subfigure}
\begin{subfigure}[b]{0.3\columnwidth}
\centering
\scalebox{0.8}{\begin{tikzpicture}[DP]
    \node[dp={1}{1}] (f) {$d$};
    \node[dp={1}{1}, below=0.3cm of f] (g) {$e$};
    \draw[runconn, runame={}] (f_res1){};
    \draw[runconn, runame={}] (g_res1){};
    \draw[funconn, funame={}] (f_fun1){};
    \draw[funconn, funame={}] (g_fun1){};
\end{tikzpicture}}
\subcaption{Parallel.}
\end{subfigure}
\begin{subfigure}[b]{0.3\columnwidth}
\centering
\scalebox{0.8}{\begin{tikzpicture}[DP]
    \node[dp={2}{2}] (f) {$d$};
    \draw[runconn, runame={}] (f_res2){};
    \draw[funconn, funame={}] (f_fun2){};
    \draw[rconn,rcname={},fcname={},feedback=1,loos=3] (f_res1) -- ($(f)+(0,5)$) |- (f_fun1);
\end{tikzpicture}}
\subcaption{Loop.}
\end{subfigure}
\label{fig:diagrams}
\caption{\glspl{abk:mdpi} can be composed in different ways.}
\label{fig:dp_def}
\end{center}
\vspace{-5mm}
\end{figure}

In this section, we present the main concepts related to the mathematical theory of co-design, presented in~\cite{Censi2015, Censi2017b} and extensively in~\cite{censi2022}.
We will recall basic concepts of order theory along the way.
For more details, the interested reader is referred to~\cite{DaveyPriestley2002}. 

The plot of the section is the following.
We will introduce a theory of co-design whose atoms are \glspl{abk:mdpi}, which are (monotone) feasibility relations between functionalities and resources.
We will then explain how one can compose \glspl{abk:mdpi} in various ways, define optimization problems related to the introduced structures, and hint at the solution techniques (with details reported in the \cref{sec:orders_background} for convenience).

Resources in this theory are quantified via partially orderes sets (posets).

\begin{definition}[Poset]
A \emph{\gls{abk:poset}} is a tuple~$\mathcal{P}=\tup{\posA,\preceq_\mathcal{P}}$, where~$\posA$ is a set and~$\preceq_\posA$ is a partial order, defined as a reflexive, transitive, and antisymmetric relation.
\end{definition}
This structure allows one to describe standard engineering quantities (typically totally ordered sets), such as $\tup{\mathbb{R}_{\geq 0},\leq}$ (e.g., energy, costs) and $\tup{\mathbb{N},\leq}$ (e.g., number of vehicles in a fleet), but also more complex ones, which we will introduce later in this work.

Given a poset, we can consider its ``reversed'' version.
\begin{definition}[Opposite of a poset]
The \emph{opposite} of a poset~$\mathcal{P}=\tup{ \posA,\preceq_\posA}$ is the poset $\tup{P,\preceq_{\mathcal{P}}^\mathrm{op}}$, which has the same elements as~$\mathcal{P}$, and the reverse ordering.
\end{definition}

To be able to define \glspl{abk:mdpi}, we further need to introduce the notions of product poset and of monotone map.

\begin{definition}[Product poset]
Let~$\tup{P,\preceq_{\mathcal{P}}}$ and~$\tup{Q,\preceq_{\mathcal{Q}}}$ be posets. Then,~$\mathcal{P}\times \mathcal{Q}=\tup{P\times Q,\preceq_{\mathcal{P}\times \mathcal{Q}}}$, with
\begin{equation*}
    \tup{p_1,q_1}\preceq_{\mathcal{P}\times \mathcal{Q}}\tup{p_2,q_2} \Leftrightarrow p_1\preceq_{\mathcal{P}}p_2 \text{ and } q_1\preceq_\mathcal{Q} q_2,
\end{equation*}
for all $p_1,p_2\in \mathcal{P}$, $q_1,q_2\in \mathcal{Q}$, is the \emph{product poset} of~$\tup{P,\preceq_{\mathcal{P}}}$ and~$\tup{Q,\preceq_{\mathcal{Q}}}$.
\end{definition}

\begin{definition}[Monotone map]
A map $f:\mathcal{P}\rightarrow \mathcal{Q}$ between two posets $\langle \posA, \preceq_\mathcal{P} \rangle$, $\langle \posB, \preceq_\mathcal{Q} \rangle$ is \emph{monotone} iff $x\preceq_\mathcal{P} y$ implies $f(x) \preceq_\mathcal{Q} f(y)$. Note that monotonicity is compositional.
\end{definition}

We are now ready to define the main atom of the monotone co-design theory.

\begin{definition}[\gls{abk:mdpi}]
Given the \glspl{abk:poset}~$\setOfFunctionalities{},\setOfResources{}$, representing \F{functionalities} and \R{resources}, respectively, we define a \emph{\glsfirst{abk:mdpi}} as a tuple~$\tup{\setOfImplementations{d},\prov, \req}$, where~$\setOfImplementations{d}$ is the set of implementations, and~$\prov$,~$\req$ are functions from~$\setOfImplementations{d}$ to~$\setOfFunctionalities{}$ and~$\setOfResources{}$, respectively:
\begin{equation*}
        \setOfFunctionalities{} \xleftarrow{\prov} \setOfImplementations{d} \xrightarrow{\req} \setOfResources{}.
\end{equation*}
\bischanged{(Maps~$\prov$,~$\req$ are mnemonics for the fact that each implementation \emph{provides} some functionality and \emph{requires} some resources.)}
We compactly denote the \gls{abk:mdpi} as~$d\colon \setOfFunctionalities{} \tickar \setOfResources{}$.
Furthermore, to each \gls{abk:mdpi} we associate a monotone map~$\bar{d}$, given by:
\begin{equation*}
    \begin{split}
        \bar{d}\colon \setOfFunctionalitiesOp{} \times \setOfResources{} &\to \tup{\power{\setOfImplementations{d}},\subseteq}\\
        \langle \F{f}^*,\R{r}\rangle &\mapsto \{i \in \setOfImplementations{d}\colon (\prov(i) \succeq_{\setOfFunctionalities{}}\F{f}) \wedge (\req(i)\preceq_{\setOfResources{}}\R{r})\},
    \end{split}
\end{equation*}
where~$(\cdot)\op$ reverses the order of a \gls{abk:poset}. We represent a \gls{abk:mdpi} in diagrammatic form as in \cref{fig:mathcodesign}.
\end{definition}

\begin{remark}[Intended semantics for \glspl{abk:mdpi}]
The expression~$\bar{d}(\F{f}^*,\R{r})$ returns the set of implementations (design choices)~$S\subseteq \setOfImplementations{d}$ for which the functionalities~$\F{f}$ are feasible with resources~$\R{r}$.
For instance, a battery provides \F{energy}, requires \R{mass} and has a \R{cost}. Different battery models represent different implementations.
\end{remark}
\begin{remark}[Monotonicity of \glspl{abk:mdpi}]
Consider an \gls{abk:mdpi} with~$\bar{d}(\F{f}^*,\R{r})=S$.
\begin{itemize}
    \item Consider~$\F{f'}\preceq \F{f}$. Then,~$\bar{d}(\F{f'}^*,\R{r})=S'\supseteq S$. In other words, decreasing the desired functionality cannot increase the required resources.
    \item Consider~$\R{r'}\succeq \R{r}$. Then,~$\bar{d}(\F{f}^*,\R{r'})=S''\supseteq S$. In other words, increasing the available resources cannot decrease the provided functionalities.
\end{itemize}
For further related examples, refer to~\cite{censi2022}.
\end{remark}

Individual \glspl{abk:mdpi} can be composed in several ways to form a co-design problem (\cref{fig:dp_def}), allowing one to decompose a large problem into smaller subproblems, and to interconnect them.
We report technical details in the appendix, and give a practical intuition in the following.
Series composition represents the case in which the functionality of a \gls{abk:mdpi} is required by another \gls{abk:mdpi}. 
For instance, the energy provided by a battery is required by an electric motor to produce torque. The posetal relation~``$\preceq$'' in \cref{fig:dp_def} represents a co-design constraint: The resource one component requires can be at most as much as the one provided by another component. 
Parallel composition formalizes decoupled processes happening together. Finally, loop composition describes feedback.\footnote{It can be proved that the formalization of feedback makes the category of \glspl{abk:mdpi} a traced monoidal category~\cite{fong2019,censi2022}.}
The composition operations preserve monotonicity, meaning that the composition of two \glspl{abk:mdpi} is a \gls{abk:mdpi}.
We call the composition of \glspl{abk:mdpi} a \gls{abk:cdpi}~\cite{Censi2015, Censi2017}.

We can now formulate problems and describe solutions.
Given a \gls{abk:poset}, we formalize the idea of ``Pareto front'' via antichains, which are useful to describe incomparable designs.
\begin{definition}[Antichains]
A subset~$S\subseteq \mathcal{P}$ of a poset $\mathcal{P}$ is an \emph{antichain} iff
no elements are comparable: For~$x,y\in S$,~$x\preceq_\mathcal{P} y$ implies $x=y$.	We denote by~$\mathsf{A}\mathcal{P}$ the set of all antichains in~$\mathcal{P}$.
\end{definition}

\begin{definition}[Functionality to resources map]
Given an~\gls{abk:mdpi}~$d$, one can define a monotone map~$h_d\colon \setOfFunctionalities{}\to \mathsf{A}\setOfResources{}$, mapping a functionality to the \emph{minimum} antichain of resources providing it. 
Dually, one can define~$h_d'\colon \setOfResources{}\to \mathsf{A}\setOfFunctionalities{}$, mapping a resource to the maximum antichain of functionalities provided by it.
\end{definition}
In the design of battery models, $h_d$ maps a particular desired energy to the antichain of masses and costs of (incomparable) battery models providing at least that energy.

We are now ready to state the problem.

\begin{problem}
\label{prob:codesign}
We are given a \gls{abk:cdpi} (interconnection of \glspl{abk:mdpi}) with functionalities $\setOfFunctionalities{}$ and resources $\setOfResources{}$, and we can evaluate the map $h_d$ for each \gls{abk:mdpi} $d$ involved. 
Given a functionality $\F{f}\in \setOfFunctionalities{}$ of the \gls{abk:cdpi}, we wish to find the \emph{minimal} resources in $\setOfResources{}$, for which there exists a feasible implementation vector which makes all sub-problems feasible simultaneously, and all co-design constraints satisfied; or, if none exist, we want to provide a certificate of infeasibility.
\end{problem}
In other words, given the maps $h_d$ for the subproblems, we want to evaluate the map $h$ for the entire \gls{abk:cdpi}.

If $h$ is Scott continuous, and the posets are complete partial orders, one can rely on Kleene's fixed point theorem to find the solution for the interconnected optimization problem~\cite{Censi2017b}.
\cref{prob:codesign} as an instance of a multi-objective optimization problem, which does not require the objectives to be convex, differentiable, continuous, or even defined on continuous spaces. 
This class of problems can be solved recursively, and their complexity is linear in the number of design options, not combinatorial.
These computational properties, and the ease of modeling within the proposed framework, make \glspl{abk:cdpi} a class of hierarchical multi-objective optimization problems that is computationally scalable, modular, and easily composable.
For convenience, we report a detailed description of the solution techniques and their complexity in \cref{sec:orders_background}.

\begin{remark}[A user-friendly framework]
The theory presented in this section and in \cref{sec:orders_background} might seem complex. However, it represents the developer-view, as opposed to the optimization framework's user-view.
As a user, one just needs to decompose the problem at hand in smaller problems, formulate them as \glspl{abk:mdpi} (via analytical relations, catalogues, and simulations), and formalize their interconnections. (A user can also focus on a single \gls{abk:mdpi} and interconnect it with other users' ones).
We will show how, once the problem is formulated, a ready-to-use solver will smoothly provide the solutions to custom queries.
\end{remark}

}
\section{Co-Design of Future Mobility Systems}
\label{sec:codesignav}
In this section, we detail our co-design framework for future mobility systems, and instantiate it for the specific case of an intermodal transportation network.

\subsection{Intermodal Mobility Framework}
\label{subsec:iamod}
\subsubsection{Multi-Commodity Flow Model}
\begin{figure}[tb]
	\begin{center}
		\includegraphics[width=\linewidth]{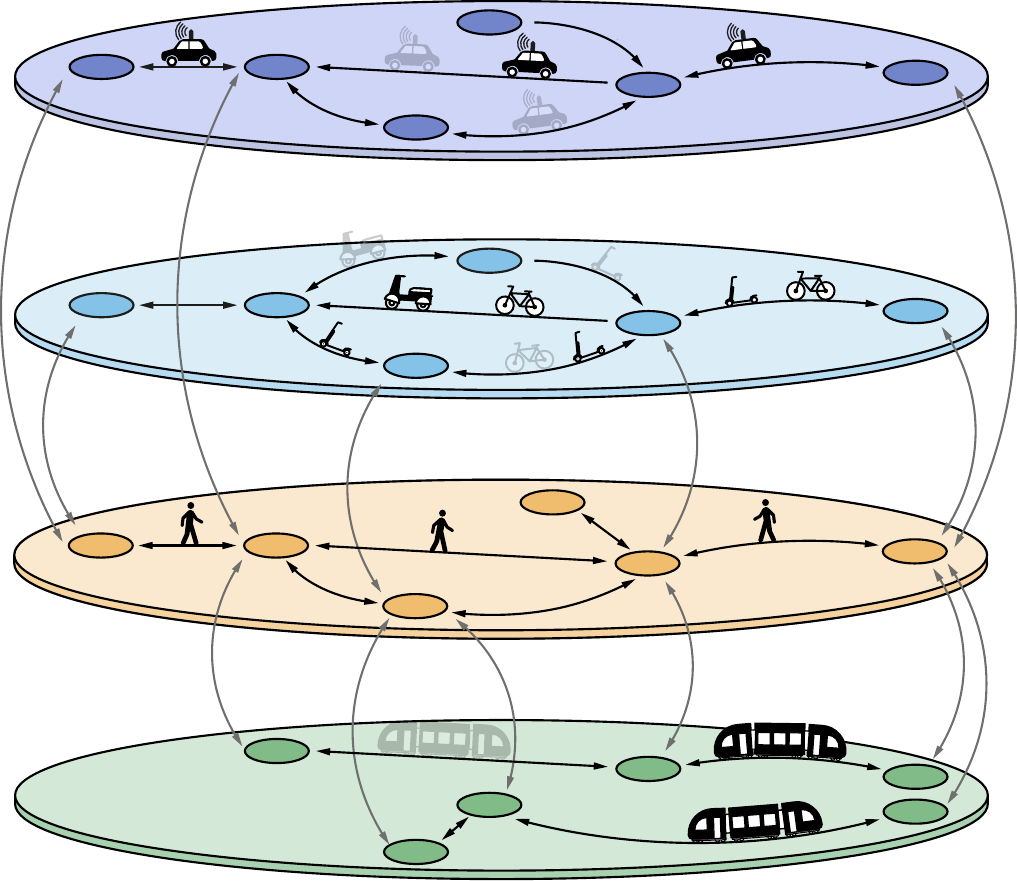}
		\caption{The intermodal AMoD network consists of road (\glspl{abk:av} and \glspl{abk:mmveh}), public transportation, and walking digraphs. The labeled circles represent stops or intersections and the black arrows denote road links, public transit arcs, or pedestrian pathways. The grey arrows represent the mode-switching arcs connecting them.}
		\label{fig:iamod}
	\end{center}
\end{figure}

The transportation system and its different modes are modeled using the edge-labeled digraph~$\cG=\tup{\setOfVertices, \setOfArcs,\colorEdge}$, sketched in Figure \ref{fig:iamod}.
It is described through a set of nodes~$\setOfVertices$ and a set of arcs~$\setOfArcs \subseteq \setOfVertices \times \setOfVertices$, labeled with metrics $\colorEdge:\setOfArcs\to\colorMetrics$. Specifically, $\colorEdge_{ij}\coloneqq \colorEdge\tup{i,j}\in\colorMetrics$ are the metrics associated to arc $\tup{i,j}\in\setOfArcs$.
Metrics of interest include edge length, travel time, energy consumption properties, and congestion models.
$\cG$ is composed of four layers: The road network layer~$\GraphRoad=\tup{\setOfVerticesRoad, \setOfArcsRoad,\colorEdgeRoad}$, consisting of an \glspl{abk:av} layer~$\GraphRoadVeh=\tup{\setOfVerticesRoadVeh, \setOfArcsRoadVeh, \colorEdgeRoadVeh}$ and a \glspl{abk:mmveh} layer~$\GraphRoadSco=\tup{\setOfVerticesRoadSco, \setOfArcsRoadSco, \colorEdgeRoadSco}$, the public transportation layer  $\GraphSubway=\tup{\setOfVerticesSubway, \setOfArcsSubway, \colorEdgeSubway}$, and a walking layer~$\GraphPedestrian=\tup{\setOfVerticesPedestrian, \setOfArcsPedestrian, \colorEdgePedestrian}$.
The \glspl{abk:av} and the \glspl{abk:mmveh} networks are characterized by intersections~$i \in \setOfVerticesRoadVeh$,~$i\in \setOfVerticesRoadSco$ and road segments~$\tup{i,j}\in \setOfArcsRoadVeh$,~$\arc \in \setOfArcsRoadSco$, respectively. Similarly, public transportation lines are modeled through station nodes~$i\in \setOfVerticesSubway$ and line segments~$\arc \in \setOfArcsSubway$.
The walking network describes walkable streets~$\arc \in \setOfArcsPedestrian$, connecting intersections~$i\in \setOfVerticesPedestrian$.
Mode-switching arcs are modeled as~
\begin{equation*}
    \begin{aligned}
    \setOfArcsCommute \subseteq &\setOfVerticesRoadVeh \times \setOfVerticesPedestrian \cup \setOfVerticesPedestrian \times \setOfVerticesRoadVeh \cup \setOfVerticesRoadSco \times \setOfVerticesPedestrian \cup \setOfVerticesPedestrian \times\\
    &\setOfVerticesRoadSco \cup \setOfVerticesSubway \times \setOfVerticesPedestrian \cup \setOfVerticesPedestrian \times \setOfVerticesSubway,
    \end{aligned}
\end{equation*}

\noindent connecting the \glspl{abk:av}, the \glspl{abk:mmveh}, and the public transportation layers to the walking layer.
To these arcs, we associate metrics $\colorEdgeConnections$.

Consequently, $\setOfVertices=\setOfVerticesPedestrian \cup \setOfVerticesRoadVeh \cup \setOfVerticesRoadSco \cup  \setOfVerticesSubway$ and $\setOfArcs=\setOfArcsPedestrian \cup \setOfArcsRoadVeh \cup \setOfArcsRoadSco \cup \setOfArcsSubway \cup \setOfArcsCommute$. Consistently with structural properties of transportation networks in urban environments, we assume $\cG$ to be strongly connected.

We now characterize the partial order of edge labeled graphs, which will be instrumental when modeling the \gls{abk:mdpi} of the mobility system. To do so, we first need to define a partial order on functions.

\begin{definition}[Poset of functions]
\label{def:posetfunctions}
Consider \glspl{abk:poset}~$A,B$, and consider the set of functions~$A\to B$, denoted by~$B^A$. 
Given any two functions~$f,g\colon A\to B$, we define
\begin{equation*}
    f\preceq_{B^A}g\Leftrightarrow f(a)\preceq_B G(a), \quad \forall a\in A.
\end{equation*}
\end{definition}
\begin{lemma}
\label{lem:funposet}
Definition~\ref{def:posetfunctions} defines a \gls{abk:poset}.
\end{lemma}
Definition~\ref{def:posetfunctions} allows one to define a poset of edge-labeled multigraphs.
\begin{definition}[Poset of edge-labeled multigraphs]
\label{def:graphposet}
Consider the set of edge-labeled multigraphs, denoted by~$\mathbf{G}$. Given~$\multigraph_1,\multigraph_2\in \mathbf{G}$, with~$\multigraph_1=\tup{\setOfVertices_1,\setOfArcs_1,c_1}$,~$\multigraph_2=\tup{\setOfVertices_2,\setOfArcs_2,c_2}$,~$c_1\colon \setOfArcs_1\to C$,~$c_2\colon \setOfArcs_2\to C$ we define the order:
\begin{equation*}
    \multigraph_1\preceq_\mathbf{G}\multigraph_2 \Leftrightarrow (\setOfVertices_1\subseteq \setOfVertices_2)\wedge  (\setOfArcs_1\subseteq \setOfArcs_2)\wedge (c_1\succeq_{C^{\setOfArcs_1}}c_2|_{\setOfArcs_1}),
\end{equation*}
where~$c_2|_{\setOfArcs_1}$ is the restriction of~$c_2$ onto the domain of~$c_1$.
\end{definition}
Intuitively, a labeled multigraph dominates another if it includes its nodes and edges, and if the labels are dominating.
\begin{lemma}
\label{lem:graphposet}
Definition \ref{def:graphposet} defines a \gls{abk:poset}.
\end{lemma}

We represent customer movements by means of travel requests. A travel request refers to a customer flow starting its trip at a node $o\in\setOfVertices$ and ending it at a node $d\in\setOfVertices$.
\begin{definition}[\bischanged{Travel demand}]
\nomenclature[a110]{$\alpha$}{Demand rate}
\nomenclature[a111]{$\alpha_\mathrm{tot}$}{Total demand rate}
A \emph{\bischanged{travel demand}}~$q$ is a triple~$\tup{o,d,\alpha} \in \setOfVertices \times \setOfVertices \times \mathbb{R}_+$, described by an origin node $o\in \setOfVertices$, a destination node $d \in \setOfVertices$, and the request rate~$\alpha >0$ (i.e., the number of customers who want to travel from~$o$ to~$d$ per unit time).
\end{definition}
Without loss of generality, we can assume that in a set of requests origin-destination pairs are not repeated, and denote the set of all possible set of requests ~$\mathcal{Q}\subset\power{\setOfVertices\times\setOfVertices\times\rplusbar}$. 
This set can be ordered as follows. 
\begin{definition}[\bischanged{Poset of travel demand}]
\label{def:poset_travel}
Consider the set of sets of travel requests~$\mathcal{Q}$. Given any~$Q_1,Q_2\in \mathcal{Q}$, one has:
\begin{equation*}
    Q_1\coloneqq\{\tup{o^1_i,d^1_i,\alpha^1_i}\}_{i=1}^{M_1}\preceq_{\mathcal{Q}}\{\tup{o^2_i,d^2_i,\alpha^2_i}\}_{i=1}^{M_2}\eqqcolon Q_2
\end{equation*}
iff for all~$\tup{o^1,d^1,\alpha^1}\in Q_1$ there is some~$\tup{o^2,d^2,\alpha^2}\in Q_2$ with~$o^1=o^2$,~$d^1=d^2$, and~$\alpha^2_i\geq \alpha^1_i$. 
In other words,~$Q_1\preceq_{\mathcal{Q}} Q_2$ if every travel request in~$Q_1$ is in~$Q_2$ as well.
\end{definition}

\begin{lemma}
\label{lem:poset_travel_isposet}
Definition~\ref{def:poset_travel} defines a poset.
\end{lemma}

To ensure that a customer is not biased to use a given transportation mode, we assume all requests to appear on the walking digraph, i.e.,  $o_m,d_m \in \setOfVerticesPedestrian $ for all $m \in \cM \coloneqq\{1,\ldots,M\}$.
The flow $\flow{i}{j}\geq 0$ describes the number of customers per unit time traveling on arc $\arc \in \setOfArcs$ and satisfying a travel request $m$. Furthermore, $\flowRebaVeh{i}{j}\geq 0$ and $\flowRebaSco{i}{j}\geq 0$ denote the flow of empty \glspl{abk:av} and \glspl{abk:mmveh} on \glspl{abk:av} arcs $\arc \in \setOfArcsRoadVeh$ and \glspl{abk:mmveh} arcs $\arc \in \setOfArcsRoadSco$, respectively. This accounts for rebalancing flows of \glspl{abk:av} and \glspl{abk:mmveh} between a customer's drop-off and the next customer's pick-up. Assuming \glspl{abk:av} and \glspl{abk:mmveh} to carry one customer at a time, the flows satisfy
\par\nobreak\vspace{-5pt}
\begingroup
\allowdisplaybreaks
\begin{small}
	\begin{subequations}
	\label{eq:flowconstotal}
		\label{eq:IAMoDConservationNonNeg}
		\begin{align}
		&\sum_{i:\arc\in\setOfArcs}\flow{i}{j} + \bool{j=o_m}\cdot \alpha_m = \sum_{k:\tup{j,k}\in\setOfArcs}\flow{j}{k} + \bool{j=d_m}\cdot \alpha_m\span\span\nonumber\\
		&  \hspace{4.8cm} \forall m\in\cM,\, j\in\cV \label{eq:flowconsa}\\
		&\sum_{i:\arc\in\setOfArcsRoadVeh} \flowTotVeh{i}{j}=\sum_{k:\tup{j,k}\in\setOfArcsRoadVeh} \flowTotVeh{j}{k}\quad \forall j\in \setOfArcsRoadVeh \label{eq:flowconsb}\\
		&\sum_{i:\arc\in\setOfArcsRoadSco} \flowTotSco{i}{j}=\sum_{k:\tup{j,k}\in\setOfArcsRoadSco} \flowTotSco{j}{k}\quad \forall j\in \setOfArcsRoadSco, \label{eq:flowconsc}
		\end{align}
	\end{subequations}
\end{small}%
\endgroup
where~$\mathbb{1}_{j=x}$ denotes the boolean indicator function,~$\flowTotVeh{i}{j}\coloneqq \flowRebaVeh{i}{j}+\sum_{m\in \mathcal{M}}f_m(i,j)$, and~$\flowTotSco{i}{j}\coloneqq \flowRebaSco{i}{j}+\sum_{m\in \mathcal{M}}f_m(i,j)$. Specifically, \eqref{eq:flowconsa} guarantees flows conservation for every transportation demand, \eqref{eq:flowconsb} preserves flow conservation for \glspl{abk:av} on every road node, and \eqref{eq:flowconsc} preserves flow conservation for \glspl{abk:mmveh} on every road node.
Combining conservation of customers~\eqref{eq:flowconsa} with the conservation of \glspl{abk:av}~\eqref{eq:flowconsb} and \glspl{abk:mmveh}~\eqref{eq:flowconsc} guarantees rebalancing \glspl{abk:av} and \glspl{abk:mmveh} to match the demand.

\subsection{Labeling Graphs with Relevant Attributes}
\label{sec:traveltime}
In the following, we specify how to label  the graphs composing the full mobility network. 
Specifically, edge-labeling maps will be of the form~$c\colon \setOfArcs \to \colorMetrics$, where~$\colorMetrics=\posreals^5$ represents link length, time needed to traverse it, its speed limit, related emissions, and capacity.
\subsubsection{Walking arcs}
We infer arc lengths~$\arcLength$ from geographical data and, assuming constant walking speed~$\arcSpeedWalking$, travel time results from~$\arcTime=\arcLength/\arcSpeedWalking.$
As speeds limits, congestion, and energy consumption do not apply to walking graphs, we set~$\arcSpeedLimit=\infty, \arcNormalCapacity=\infty$,~$\arcEnergy=0$.
Accordingly, 
\begin{equation*}
\begin{aligned}
    \colorEdgePedestrian:
    \setOfArcsPedestrian &\to \colorMetrics
    \\
    \arc &\mapsto \tup{\arcLength,\arcTime,\arcSpeedLimit,\arcEnergy,\arcNormalCapacity}.
\end{aligned}
\end{equation*}

\subsubsection{Public transit arcs}
We infer arc lengths $\arcLength$ from public transit network data.
\changed{Furthermore, assuming that the public transportation system at node $j$ operates with the frequency $\freqTrain$, travel time results from~$\arcTime=\arcTime^\mathrm{nom}+\timePedestrianSubway+ 1/(2\freqTrain)$,
where~$\arcTime^\mathrm{nom}$ is the in-vehicle travel time (inferred from public transit schedules) and~$\timePedestrianSubway$ is a constant sidewalk-to-station travel time.}
We ignore capacity and speed limits, so that~$\arcNormalCapacity=\arcSpeedLimit=\infty$.
For the public transportation system we assume a constant energy consumption per unit time. This approximation is reasonable in urban environments, where the operation of the public transportation system is independent from the number of customer serviced, and its energy consumption is therefore invariant. Therefore, we write~$\arcEnergy=\kappa \arcTime$,~$\kappa>0$.
Accordingly, 
\begin{equation*}
\begin{aligned}
    \colorEdgeSubway\colon
    \setOfArcsSubway &\to \colorMetrics
    \\
    \arc &\mapsto \tup{\arcLength,\arcTime,\arcSpeedLimit,\arcEnergy,\arcNormalCapacity}.
\end{aligned}
\end{equation*}

\subsubsection{Road arcs for \glspl{abk:av}}
Each road arc is characterized by a length~$\arcLength$, a speed limit~$\arcSpeedLimit$, and a capacity~$\arcNormalCapacity$, all derived from road network data.
We consider \glspl{abk:av} driving at speed~$\achievableSpeedVeh$, so that travel time reads
\begin{equation*}
    \arcTime = \frac{\arcLength}{\min\{\achievableSpeedVeh, \arcSpeedLimit\}}.
\end{equation*}
We compute the energy consumption of \glspl{abk:av} via an urban driving cycle. In particular, the cycle is scaled so that its average speed~$v_\mathrm{avg,cycle}$ matches the free-flow speed on the link. The energy consumption of road link~$\arc$ is scaled as
\begin{equation*}
\arcEnergy=e_\mathrm{cycle} \cdot \frac{\arcLength}{s_\mathrm{cycle}}.
\end{equation*}
Collectively, 
\begin{equation*}
\begin{aligned}
    \colorEdgeRoadVeh\colon
    \setOfArcsRoadVeh &\to \colorMetrics
    \\
    \arc &\mapsto \tup{\arcLength,\arcTime,\arcSpeedLimit,\arcEnergy,\arcNormalCapacity}.
\end{aligned}
\end{equation*}

\subsubsection{Road arcs for \glspl{abk:mmveh}}
Each road arc is characterized by a length $\arcLength$ and a speed limit $\arcSpeedLimit$, derived from road network data, while we neglect arc capacity (i.e., $\arcNormalCapacity=\infty$).
Assuming \glspl{abk:mmveh} driving at speed~$\achievableSpeedSco$, travel time reads
\begin{equation*}
\begin{aligned}
    \arcTime=\frac{\arcLength}{\min\{\achievableSpeedSco,\arcSpeedLimit\}}.
\end{aligned}
\end{equation*}
For \glspl{abk:mmveh} we consider a distance-based energy consumption, i.e.~$\arcEnergy=\iota \arcLength$, with~$\iota >0$.
Overall, 
\begin{equation*}
\begin{aligned}
    \colorEdgeRoadSco\colon
    \setOfArcsRoadSco &\to \colorMetrics
    \\
    \arc &\mapsto \tup{\arcLength,\arcTime,\arcSpeedLimit,\arcEnergy,\arcNormalCapacity}.
\end{aligned}
\end{equation*}

\subsubsection{\bischanged{Transfer} arcs}
We define travel time $\arcTime$ as follows: we assume that the average waiting time for \glspl{abk:av} is $t_\mathrm{WV}$, the average time needed to reach a \gls{abk:mmveh} is $\timePedestrianRoadSco$, and switching from the \glspl{abk:av} graph, the \glspl{abk:mmveh} graph, and the public transit graph to the pedestrian graph takes the transfer times $\timeRoadVehPedestrian$, $\timeRoadScoPedestrian$, and $\timeSubwayPedestrian$, respectively. 
For each arc, we set length and energy consumption to zero (i.e., $\arcLength=\arcEnergy=0$) and ignore capacity and speed limit (i.e., $\arcNormalCapacity=\arcSpeedLimit=\infty$). Overall, 
\begin{equation*}
\begin{aligned}
    \colorEdgeConnections\colon
    \setOfArcsCommute &\to \colorMetrics
    \\
    \arc &\mapsto \tup{\arcLength,\arcTime,\arcSpeedLimit,\arcEnergy,\arcNormalCapacity}.
\end{aligned}
\end{equation*}

\subsection{Road Congestion}
We assume that road arcs are subject to a normalized capacity~$\arcNormalCapacity$, which could arise from the difference of the nominal road capacity
and the exogenous road usage:
\begin{equation}
\label{eq:capacity}
\flowTotVeh{i}{j} \leq \arcNormalCapacity.
\end{equation}
We assume that the central authority operates the \gls{abk:amod} fleet such that vehicles travel at free-flow speed throughout the road network of the city, meaning that the total flow on each road link must be below the link's capacity.
Therefore, we capture congestion effects with the threshold model.
Finally, we assume \gls{abk:mm} to not significantly contribute to congestion~\cite{Blackwell2019}.

\subsection{Discussion}
First, the demand is assumed to be time-invariant and flows are allowed to have fractional values. This assumption is in line with the mesoscopic and system-level planning perspective of our study.
Second, we model congestion effects using a threshold model. This approach can be interpreted as a municipality preventing mobility solutions to exceed the critical flow density on road arcs. \glspl{abk:av} and \glspl{abk:mmveh} can therefore be assumed to travel at free flow speed~\cite{Daganzo2008}. This assumption is realistic for an initial low penetration of new mobility systems in the transportation market, especially when the \gls{abk:av} and \gls{abk:mmveh} fleets are limited in size.
Finally, we allow \glspl{abk:av} and \glspl{abk:mmveh} to transport one customer at a time~\cite{Henao2019}.

\subsection{Co-Design Framework}
\label{sec:codesignframework}

We integrate the intermodal framework presented in \cref{subsec:iamod} in the co-design formalism, allowing the decoupling of the \gls{abk:cdpi} of a complex system in the \glspl{abk:mdpi} of its individual components in a modular, compositional, and systematic fashion. 
To achieve this, we decouple the \gls{abk:cdpi} in the \glspl{abk:mdpi} of the individual \gls{abk:av} (\cref{sec:vehdp}), the \glspl{abk:av} fleet (\cref{sec:iamod_codesign_1,sec:iamod_codesign_2,sec:iamod_codesign_3}), the individual \gls{abk:mmveh} (\cref{sec:escooterdp}), the \glspl{abk:mmveh} fleet (\cref{sec:iamod_codesign_1,sec:iamod_codesign_2,sec:iamod_codesign_3}), and the public transportation system (\cref{sec:subdp}). Their interconnection is presented in \cref{sec:mcdp_1,sec:mcdp_2,sec:mcdp_3}, where we propose multiple model versions, showcasing the flexibility of the developed framework.
We aim at computing the antichain of resources, quantified in terms of costs, average travel time per trip, and emissions required to provide the mobility service to a set of customers.
For each model, we provide descriptions and formal proofs of integration in the co-design framework.

\subsubsection{The \gls{abk:av} \gls{abk:mdpi}}
\label{sec:vehdp}
The \gls{abk:av} \gls{abk:mdpi} selects the labeled graph on which the \gls{abk:av} provider wants to operate. The selection happens via the choice of the achievable speed of the \glspl{abk:av} as follows. 
\glspl{abk:av} safety protocols impose a maximum achievable velocity~$\achievableSpeedVeh$. Furthermore, in order to prevent too slow and therefore dangerous driving behaviors~\cite{Dahl2018}, we only consider \glspl{abk:av} arcs through which the \glspl{abk:av} can drive at least at a fraction~$\beta$ of the speed limit.\nomenclature[a102]{$\beta$}{Threshold for minimal speed}
Specifically, \glspl{abk:av} can drive on arc $\arc \in \setOfArcsRoadVeh$ if and only if 
\begin{equation}
    \label{eq:droparcs}
    \achievableSpeedVeh \geq \beta \cdot \pi_{v_\mathrm{L}}\colorEdgeRoadVeh(i,j),
\end{equation}
where~$\beta\in (0,1]$, and \bischanged{$\pi_{v_\mathrm{L}}$ projects the part of $\colorEdgeRoadVeh(i,j)$ related to $v_\mathrm{L}$.}
The elimination of forbidden arcs given an achievable speed can be achieved through the following map (\bischanged{mnemonics for reduction}):
\begin{equation*}
    \begin{aligned}
    \mathrm{red}_\mathrm{R,V}\colon \mathbb{R}_{\geq 0}&\to \tup{\setofGraphs,\preceq_{\setofGraphs}}\\
    \achievableSpeedVeh&\mapsto \tup{\setOfVerticesRoadVeh,\setOfArcs,\colorEdge},
    \end{aligned}
\end{equation*}
where
\begin{equation}
\label{eq:red_condition_road_veh}
\begin{aligned}
    \setOfArcs &=\{ a\in\setOfArcsRoadVeh: \text{\cref{eq:droparcs} holds}\},
    \\
    \colorEdge &=\tup{\pi_s\colorEdgeRoadVeh,\frac{\pi_s\colorEdgeRoadVeh}{\min\{ \achievableSpeedVeh, \pi_{v_\mathrm{L}}\colorEdgeRoadVeh\}},\pi_{v_\mathrm{L}}\colorEdgeRoadVeh,\pi_e\colorEdgeRoadVeh,\pi_k\colorEdgeRoadVeh}.
\end{aligned}
\end{equation}
\begin{lemma}
\label{lem:red_road_monotone}
    The map~$\mathrm{red}_\mathrm{R,V}$ is monotone.
\end{lemma}
Under the rationale that driving safely at higher speed requires more advanced sensing and algorithmic capabilities~\cite{zardiniIROS2020}, we model the achievable speed of the \glspl{abk:av} $\F{\achievableSpeedVeh}$ as a \emph{monotone} function of the vehicle fixed costs $\R{\costFixVeh}$ (resulting from the cost of the vehicle $C_\mathrm{V,v}$ and the cost of its automation $C_\mathrm{V,a}$) and the mileage-dependent operational costs $\R{\costOpVeh}$ (accounting for maintenance, cleaning, energy consumption, depreciation, and opportunity costs~\cite{Mas-ColellWhinstonEtAl1995}).
\paragraph*{\gls{abk:mdpi} Definition}
The \gls{abk:av} \gls{abk:mdpi}, denoted~$d_\mathrm{AV}$, provides the functionality~$\F{\cG_{\mathrm{AV}}}\in \F{\setofGraphs}$ (i.e., the functionality of servicing a specific network with a specific performance) and requires the resources $\R{\costFixVeh},\R{\costOpVeh}\in \rplusbar$.
The implementations space $\setOfImplementations{\mathrm{AV}}$ consists of models of the \glspl{abk:av}.
Formally:~$d_\mathrm{AV}\colon \F{\setofGraphs} \tickar \R{\rplusbar}^2.$
\begin{lemma}
\label{lem:av_dp_welldef}
$d_\mathrm{AV}$ is a well-defined \gls{abk:mdpi}.
\end{lemma}
\subsubsection{The \gls{abk:mmveh} \gls{abk:mdpi}}
\label{sec:escooterdp}
The \gls{abk:mm} \gls{abk:mdpi} comprises the selection of the labeled graph on which to operate, again resumed in the maximal speed achievable by \glspl{abk:mmveh}.
Given an achievable speed~$\achievableSpeedSco$, one obtains the resulting graph as follows:
\begin{equation*}
    \begin{aligned}
    \mathrm{red}_\mathrm{R,M}\colon \mathbb{R}_{\geq 0}&\to \tup{\setofGraphs,\preceq_{\setofGraphs}}\\
    \achievableSpeedSco&\mapsto \tup{\setOfVerticesRoadSco,\setOfArcsRoadSco,\colorEdge},
    \end{aligned}
\end{equation*}
where
\begin{equation*}
\begin{aligned}
    \colorEdge &=\tup{\pi_s\colorEdgeRoadSco,\frac{\pi_s\colorEdgeRoadSco}{\min\{ \achievableSpeedSco, \pi_{v_\mathrm{L}}\colorEdgeRoadSco\}},\pi_{v_\mathrm{L}}\colorEdgeRoadSco,\pi_e\colorEdgeRoadSco,\pi_k\colorEdgeRoadSco}.
\end{aligned}
\end{equation*}
\begin{lemma}
\label{lem:red_veh_micro_monotone}
    The map~$\mathrm{red}_\mathrm{R,M}$ is monotone.
\end{lemma}
Following the rationale that different \glspl{abk:mmveh} can reach different speeds and have different prices, we model the achievable speed of the \gls{abk:mmveh}~$\F{\achievableSpeedSco}$ as a \emph{monotone} function of the \gls{abk:mmveh} fixed costs~$\R{\costFixSco}$ and the mileage-dependent operational costs~$\R{\costOpSco}$.
\paragraph*{\gls{abk:mdpi} Definition}
Therefore, the \gls{abk:mm} \gls{abk:mdpi}, denoted~$d_\mathrm{MM}$, provides the functionality~$\F{\cG_{\mathrm{MM}}}\in \F{\setofGraphs}$ (i.e., the functionality of servicing a specific network with a specific performance) and requires the resources $\R{\costFixSco},\R{\costOpSco}\in \R{\rplusbar}$.
The implementations space $\setOfImplementations{\mathrm{M}}$ consists of instances of the \glspl{abk:mmveh}.
Formally:~$d_\mathrm{MM}\colon \F{\setofGraphs} \tickar \R{\rplusbar}^2.$

\begin{lemma}
\label{lem:dp_mm_welldef}
$d_\mathrm{MM}$ is a well-defined \gls{abk:mdpi}.
\end{lemma}

\subsubsection{The Subway \gls{abk:mdpi}}
\label{sec:subdp}
The public transit \gls{abk:mdpi} comprises the selection of the labeled network on which to operate, now resumed in the choice of fleet size for the subway system.
Specifically, we assume the service frequency~$\freqTrain$ to scale monotonically with the size of the train fleet~$\numberFleetTrain$. 
In the linear case, one has:
\begin{equation*}
    \frac{\freqTrain}{\freqTrainBaseline}=\frac{\numberFleetTrain}{\numberFleetTrainBaseline},
\end{equation*}
where~$\freqTrainBaseline$ and~$\numberFleetTrainBaseline$ are respective existing baselines. 
Given a train fleet size, one obtains the resulting network as follows:
\begin{equation*}
    \begin{aligned}
    \mathrm{red}_\mathrm{P}\colon \mathbb{N}&\to \tup{\setofGraphs,\preceq_{\setofGraphs}}\\
    \numberFleetTrain&\mapsto \tup{\setOfVerticesSubway,\setOfArcsSubway,\colorEdge},
    \end{aligned}
\end{equation*}
where
\begin{equation*}
\begin{aligned}
    \colorEdge &=\tup{\pi_s\colorEdgeSubway,\timePedestrianSubway+ \frac{\numberFleetTrainBaseline}{2\numberFleetTrain\freqTrainBaseline}
    ,\pi_{v_\mathrm{L}}\colorEdgeSubway,\pi_e\colorEdgeSubway,\pi_k\colorEdgeSubway}.
\end{aligned}
\end{equation*}
\begin{lemma}
\label{lem:red_pub_monotone}
    The map~$\mathrm{red}_\mathrm{P}$ is monotone.
\end{lemma}

We relate a train fleet of size~$\numberFleetTrain$ to the fixed costs $\costFixTrain$ (accounting for train and infrastructural costs) and to the operational costs~$\costOpTrain$ (accounting for energy consumption, vehicles depreciation, and train operators' wages).
Given the passengers-independent public transit operation in today's cities, we assume the operational costs $\R{\costOpTrain}$ to be mileage independent and to only vary with the size of the fleet.
Assuming an average train's life of $\lifeTrain$, and a baseline subway fleet of $n_\mathrm{S,baseline}$ trains, costs are 
\begin{equation*}
    \R{\costSub}=\frac{\costFixTrain}{\lifeTrain}\cdot n_\mathrm{S,a} + \costOpTrain.
\end{equation*}
Moreover, operating a fleet of trains entails the CO\textsubscript{2} emissions
\begin{equation*}
    \R{\emissionsSub}=\emissionsSubTrain \cdot \numberFleetTrain.
\end{equation*}

\paragraph*{\gls{abk:mdpi} Definition}
The public transit \gls{abk:mdpi}, denoted~$d_\mathrm{P}$, provides the functionality~$\F{\cG_{\mathrm{P}}}\in \F{\setofGraphs}$ (i.e., the functionality of servicing a specific network with a specific performance) and requires the resources $\R{\costSub}\in \rplusbar$ and $\R{\emissionsSub}\in \rplusbar$.
The implementations space $\setOfImplementations{\mathrm{P}}$ consists of different train acquisition choices.
Formally:~$d_\mathrm{P}\colon \F{\setofGraphs} \tickar \R{\rplusbar}^2.$
\begin{lemma}
\label{lem:dp_pub_welldef}
$d_\mathrm{P}$ is a well-defined \gls{abk:mdpi}.
\end{lemma}

\subsubsection{The Intermodal Mobility System \gls{abk:mdpi} (Version 1)}
\label{sec:iamod_codesign_1}
The first version of the intermodal mobility system \gls{abk:mdpi} considers demand satisfaction as a functionality. 

To successfully satisfy a given set of travel requests, we require the following resources:
\begin{itemize}
    \item the mobility network resulting from the design of \glspl{abk:av}, \glspl{abk:av}~$\R{\cG_{\mathrm{AV}}}=\tup{\setOfVertices_\mathrm{AV},\setOfArcs_{\mathrm{AV}}, \colorEdge_\mathrm{AV}}$, 
    \item the mobility network resulting from the design of public transit~$\R{\cG_{\mathrm{P}}}=\tup{\setOfVertices_\mathrm{P},\setOfArcs_{\mathrm{P}}, \colorEdge_\mathrm{P}}$,
    \item  the number of available \glspl{abk:av} per fleet~$\R{\numberFleetVeh}$,
    \item the average travel time of a trip
    \begin{equation*}
        \label{eq:traveltime_1}
        \R{\averageTravelTime} \coloneqq\frac{1}{\alpha_\mathrm{tot}} \sum_{\substack{m\in \mathcal{M}, \\ \arc \in \setOfArcsPedestrian \cup \setOfArcs_{\mathrm{AV}} \cup \setOfArcs_{\mathrm{P}} \cup \setOfArcsCommute}} \pi_t\colorEdge(i,j) \cdot \flow{i}{j},
    \end{equation*}
    with
    \begin{equation}
    \label{eq:alpha_tot}
    \alpha_\mathrm{tot}\coloneqq \sum_{m\in\mathcal{M}}\alpha_m,
    \end{equation}
    \item the total distance driven by the \glspl{abk:av} per unit time
    \begin{equation}
    \label{eq:distance_veh}
        \R{\distanceVeh} \coloneqq\sum_{\arc \in \setOfArcs_{\mathrm{AV}}}\pi_s\colorEdge_\mathrm{AV}(i,j) \cdot \flowTotVeh{i}{j},
    \end{equation}
    \item the total \glspl{abk:av} CO\textsubscript{2} emissions per unit time
       \begin{equation}
       \label{eq:emissions_veh}
        \R{\emissionsVeh} \coloneqq\gamma \cdot \sum_{\arc \in \setOfArcs_{\mathrm{AV}}} \pi_e\colorEdge_\mathrm{AV} \cdot \flowTotVeh{i}{j}.
    \end{equation}
\end{itemize}    
We assume that \glspl{abk:av} are routed to maximize the customers' welfare, defined without loss of generality as the average travel time~$\R{\averageTravelTime}$. 
Hence, we link functionality and resources of the mobility system \gls{abk:mdpi} through the optimization problem:
\begin{equation}
\begin{aligned}
\label{eq:TIamodopt_1}
\min_{\substack{\{f_m\}_m \\ f_{0,\mathrm{V}} }}
&\averageTravelTime\\
\mathrm{ s.t.\ }
&\mathrm{Eq. } \eqref{eq:flowconstotal},
\\
&\mathrm{ Eq. }  \eqref{eq:capacity}\quad \forall \arc\in\setOfArcs_\mathrm{AV},
\\
&\sum_{\arc \in \setOfArcs_\mathrm{AV}} \flowTotVeh{i}{j} \cdot \pi_{t}\colorEdge_\mathrm{AV}(i,j)
    \leq \numberFleetVeh,
\end{aligned}
\end{equation}
where we express the number of vehicles on arc $\arc$ as the multiplication of the total vehicles flow on the arc and its travel time.

\paragraph*{\gls{abk:mdpi} Definition}
The intermodal mobility system \gls{abk:mdpi} has as functionality the satisfied requests~$\F{Q}\in \F{\mathcal{Q}}$ and the mentioned resources.
Furthermore, $\setOfImplementations{\mathrm{O}}$ consists of specific intermodal scenarios.
Formallyn:~$d_\mathrm{IAMOD}\colon \F{\mathcal{Q}}\tickar \R{\setofGraphs}^2\times \R{\bar{\mathbb{N}}}\times \R{\rplusbar}^3.$
\begin{lemma}
\label{lem:iamod_1_dp_welldef}
$d_\mathrm{IAMOD,1}$ is a well-defined \gls{abk:mdpi}.
\end{lemma}

\subsubsection{The Intermodal Mobility System \gls{abk:mdpi} (Version 2)}
\label{sec:iamod_codesign_2}
The second version of the intermodal mobility system \gls{abk:mdpi} still considers demand satisfaction as a functionality, now including \gls{abk:mm} options.
To successfully satisfy a given set of travel requests, we require the following resources:
\begin{itemize}
    \item $\R{\cG_{\mathrm{AV}}}=\tup{\setOfVertices_\mathrm{AV},\setOfArcs_{\mathrm{AV}}, \colorEdge_\mathrm{AV}}$ as in \cref{sec:iamod_codesign_1},
    \item $\R{\cG_{\mathrm{P}}}=\tup{\setOfVertices_\mathrm{P},\setOfArcs_{\mathrm{P}}, \colorEdge_\mathrm{P}}$ as in \cref{sec:iamod_codesign_1},
    \item the mobility network resulting from the design of \glspl{abk:mm}, \glspl{abk:mm}~$\R{\cG_{\mathrm{MM}}}=\tup{\setOfVertices_\mathrm{MM},\setOfArcs_{\mathrm{MM}}, \colorEdge_\mathrm{MM}}$,
    \item $\R{\numberFleetVeh}$ as in \cref{sec:iamod_codesign_1},
    \item the number of available \glspl{abk:mmveh} per fleet $\R{\numberFleetSco}$, 
    \item the (adapted) average travel time of a trip
    \begin{equation*}
        \label{eq:traveltime_2}
        \R{\averageTravelTime} \coloneqq\frac{1}{\alpha_\mathrm{tot}} \sum_{\substack{m\in \mathcal{M}, \\ \arc \in \setOfArcsPedestrian \cup \setOfArcs_{\mathrm{AV}} \cup \setOfArcs_{\mathrm{MM}} \cup \setOfArcs_{\mathrm{P}} \cup \setOfArcsCommute}} \hspace{-1cm}\pi_t\colorEdge(i,j) \cdot \flow{i}{j},
    \end{equation*}
    with~$\alpha_\mathrm{tot}$ as in \cref{eq:alpha_tot},
    \item $\R{\distanceVeh}$ as in \cref{eq:distance_veh},
    \item the total distance driven by the \glspl{abk:mmveh} per unit time
    \begin{equation*}
        \R{\distanceSco} \coloneqq\sum_{\arc \in \setOfArcs_{\mathrm{MM}}}\pi_s\colorEdge_\mathrm{MM}(i,j) \cdot 
        \flowTotSco{i}{j},
    \end{equation*}
    \item $\R{\emissionsVeh}$ as in \cref{eq:emissions_veh},
    \item the total \glspl{abk:mmveh} CO\textsubscript{2} emissions per unit time
       \begin{equation*}
        \R{\emissionsSco} \coloneqq\gamma \cdot \sum_{\arc \in \setOfArcs_{\mathrm{MM}}} \pi_e\colorEdge_\mathrm{MM} \cdot \flowTotSco{i}{j} ,
    \end{equation*}
    where~$\gamma$ relates energy consumption and CO\textsubscript{2} emissions.\nomenclature[a120]{$\gamma$}{Energy consumption to CO\textsubscript{2} emissions}
\end{itemize}    
We assume that \glspl{abk:av} and \glspl{abk:mmveh} are routed to maximize the customers' welfare, defined without loss of generality as the average travel time~$\R{\averageTravelTime}$. 
Hence, we link functionality and resources of the mobility system \gls{abk:mdpi} through the following optimization problem, extending \cref{eq:TIamodopt_1}:
\begin{equation}
\begin{aligned}
\label{eq:TIamodopt_2}
\min_{\substack{\{f_m\}_m \\ f_{0,\mathrm{V}} \\ f_{0,\mathrm{M}}}}
&\averageTravelTime\\
\mathrm{ s.t.\ }
&\mathrm{Eq. } \eqref{eq:flowconstotal},
\\
&\mathrm{ Eq. }  \eqref{eq:capacity}\quad \forall \arc\in\setOfArcs_\mathrm{AV},
\\
&\sum_{\arc \in \setOfArcs_\mathrm{AV}} \flowTotVeh{i}{j} \cdot \pi_{t}\colorEdge_\mathrm{AV}(i,j)
    \leq \numberFleetVeh,
\\
&\sum_{\arc \in \setOfArcs_\mathrm{MM}} \flowTotSco{i}{j} \cdot \pi_{t}\colorEdge_\mathrm{MM}(i,j)
    \leq \numberFleetSco,
\end{aligned}
\end{equation}
where we express the number of vehicles on arc $\arc$ as the multiplication of the total vehicles flow on the arc and its travel time.
\paragraph*{\gls{abk:mdpi} Definition}
The intermodal mobility system \gls{abk:mdpi} has as functionality~$\F{Q}\in \F{\mathcal{Q}}$ and the mentioned resources.
Furthermore, $\setOfImplementations{\mathrm{O}}$ consists of specific intermodal scenarios.
Formally:~$d_\mathrm{IAMOD,2}\colon \F{\mathcal{Q}}\tickar \R{\setofGraphs}^3\times \R{\bar{\mathbb{N}}}^2\times \R{\rplusbar}^5.$
\begin{lemma}
\label{lem:iamod_2_dp_welldef}
$d_\mathrm{IAMOD,2}$ is a well-defined \gls{abk:mdpi}.
\end{lemma}

\subsubsection{The Intermodal Mobility System \gls{abk:mdpi} (Version 3)}
\label{sec:iamod_codesign_3}
We extend the setting presented in \cref{sec:iamod_codesign_1} by including a new functionality.
Specifically, the intermodal mobility system \gls{abk:mdpi} not only provides demand satisfaction as a functionality, but also provides the revenue~$\rho$ arising from the mobility offer, which reads:
\begin{equation*}
    \F{\rho}= p_\mathrm{AV}\distanceVeh+ p_\mathrm{P}\sum_{\arc\in\setOfArcsCommute\cap\setOfVerticesPedestrian\times\setOfVerticesSubway}\flow{i}{j},
\end{equation*}
where~$p_\mathrm{AV}$ is a distance-based price to use \glspl{abk:av} and~$p_\mathrm{P}$ is a fixed entry price for the subway system. Accordingly, we modify the optimization problem to account for both average travel time and average cost of fare: 
\begin{equation}
\begin{aligned}
\label{eq:TIamodopt_3}
\min_{\substack{\{f_m\}_m \\ f_{0,\mathrm{V}} }}
&V_\mathrm{T}\averageTravelTime + \frac{1}{\alpha_\mathrm{tot}}\rho \\
\mathrm{ s.t.\ }
&\mathrm{Eq. } \eqref{eq:flowconstotal},
\\
&\mathrm{ Eq. }  \eqref{eq:capacity}\quad \forall \arc\in\setOfArcs_\mathrm{AV},
\\
&\sum_{\arc \in \setOfArcs_\mathrm{AV}} \flowTotVeh{i}{j} \cdot \pi_{t}\colorEdge_\mathrm{AV}(i,j)
    \leq \numberFleetVeh,
\end{aligned}
\end{equation}
where $V_\mathrm{T}$ is the value of time.

\paragraph*{\gls{abk:mdpi} Definition}
This new version of the intermodal mobility system \gls{abk:mdpi} has as functionality the satisfied requests~$\F{Q}\in \F{\mathcal{Q}}$ and the total revenue~$\F{\rho}\in \F{\rplusbar}$ and the mentioned resources.
Furthermore,~$\setOfImplementations{\mathrm{O}}$ consists of specific intermodal scenarios (including specific price choices).
Formally:~$d_\mathrm{IAMOD,3}\colon \F{\mathcal{Q}}\times \F{\rplusbar}\tickar \R{\setofGraphs}^2\times \R{\bar{\mathbb{N}}}\times \R{\rplusbar}^3.$
\begin{lemma}
\label{lem:iamod_3_dp_welldef}
$d_\mathrm{IAMOD,3}$ is a well-defined \gls{abk:mdpi}.
\end{lemma}

\subsubsection{The Mobility \gls{abk:mdpi} (Version 1)}
\label{sec:mcdp_1}

\begin{figure}[tb]
    \begin{center}
    \begin{subfigure}[b]{\columnwidth}
    \includegraphics[width=\columnwidth]{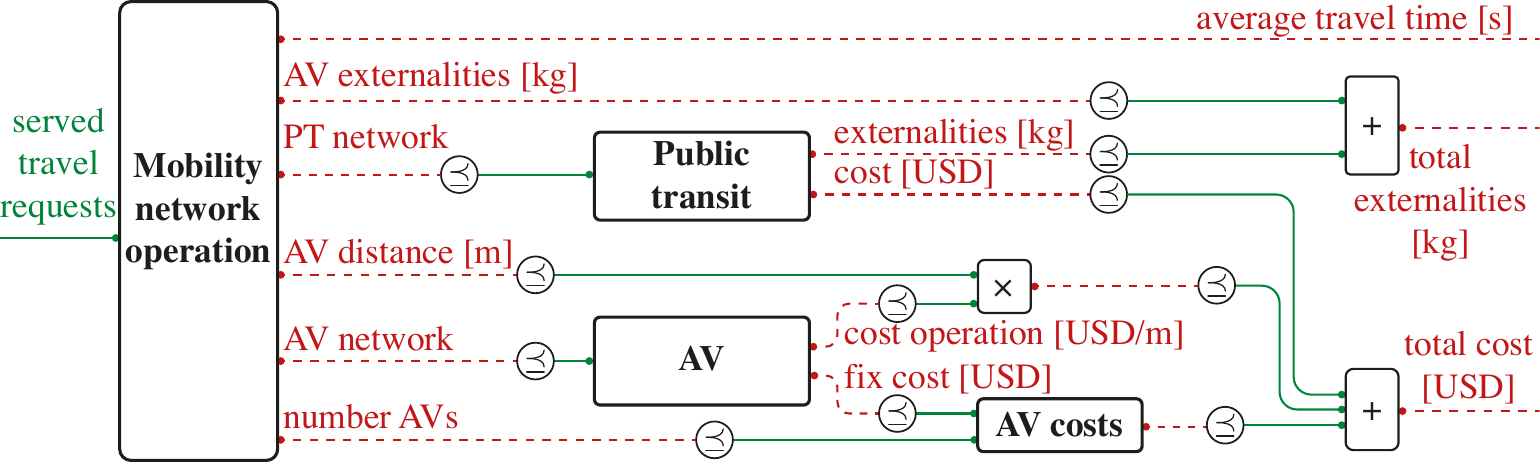}
    \caption{MDPI~$d_\mathrm{Mob_1}$.}
    \label{fig:dpmob_1}
    \end{subfigure}
    ~
    \begin{subfigure}[b]{\columnwidth}
    \includegraphics[width=\columnwidth]{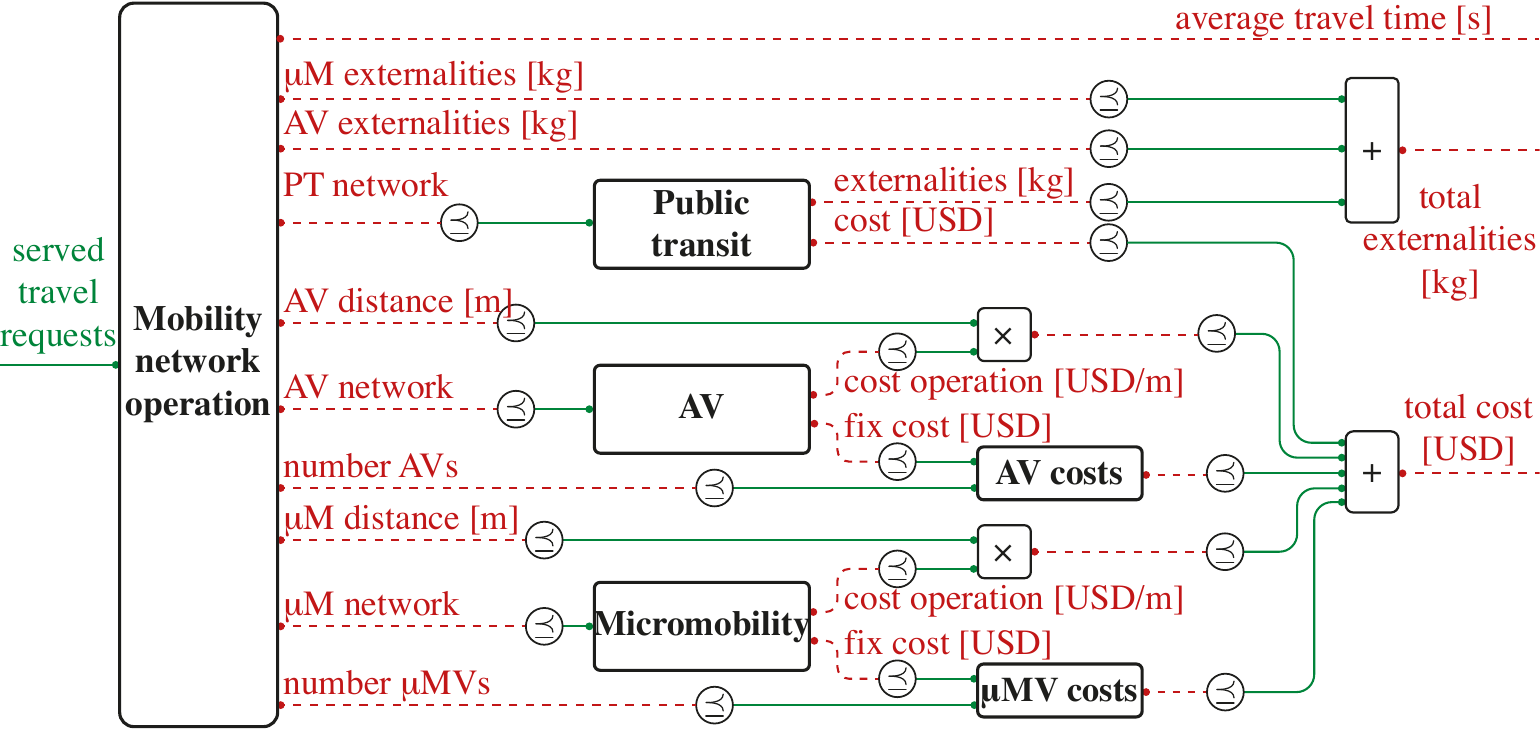}
    \caption{MDPI~$d_\mathrm{Mob_2}$.}
    \label{fig:dpmob_2}
    \end{subfigure}
    ~
    \begin{subfigure}[b]{\columnwidth}
    \includegraphics[width=1\columnwidth]{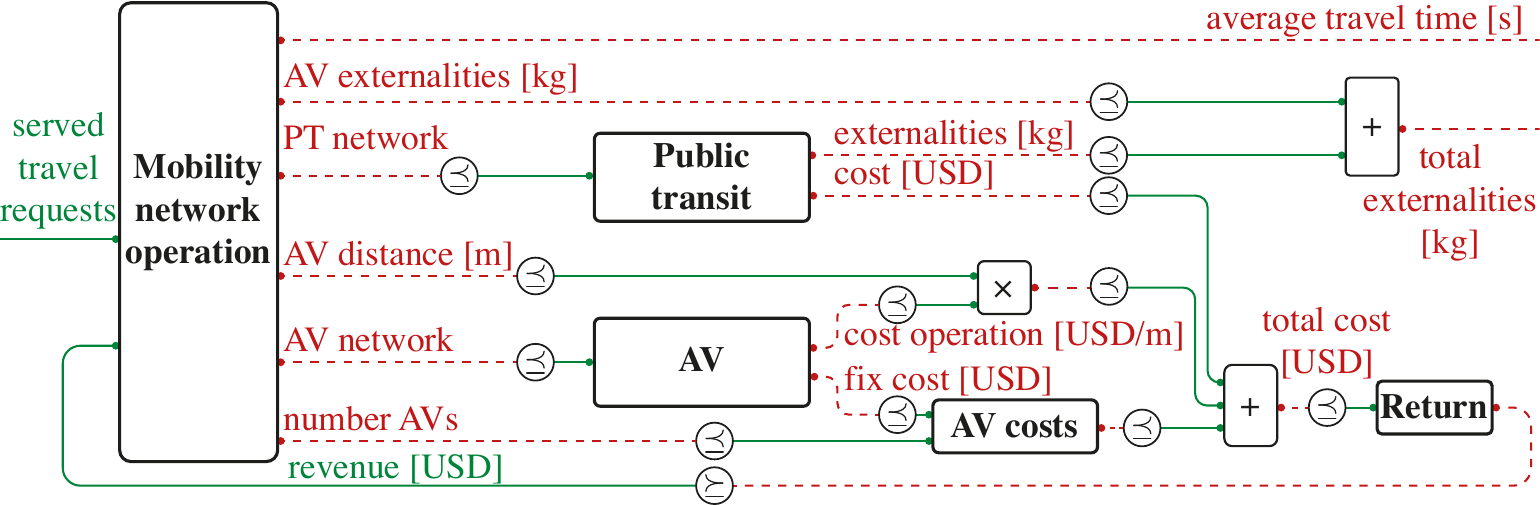}
    \caption{MDPI~$d_\mathrm{Mob_3}$.}
    \label{fig:dpmob_3}
    \end{subfigure}
    \caption{Co-design diagrams for the \glspl{abk:mdpi} defined in \cref{sec:codesignframework}.}
    \label{fig:diagrams}
    \end{center}
\end{figure}

The functionality of the system is to satisfy the customers' demand. 
Formally, the functionality provided by the \gls{abk:cdpi} is the set of travel requests and coincides with the functionalities of~$d_\mathrm{I_1}$.
To provide the mobility service, three resources are required. First, on the customers' side, we require the average travel time defined in \cref{eq:traveltime_1}.
Second, on the side of the central authority, the resource is the total transportation cost of the intermodal mobility system. 
Assuming an average \gls{abk:av}'s life of~$\lifeVeh$, an average \gls{abk:mmveh}'s life of $\lifeSco$, we express the total costs as
\begin{equation*}
\R{\costTot}=\costVeh + \costSub,
\end{equation*}
where~$\costVeh$ is the \glspl{abk:av}-related cost
\begin{equation*}
\costVeh=\frac{\costFixVeh}{\lifeVeh}\cdot \numberFleetVeh + \costOpVeh \cdot \distanceVeh,
\end{equation*}
and~$\costSub$ is the public transit-related cost.
Third, on the environmental side, we consider the total CO\textsubscript{2} emissions 
\begin{equation*}
\R{\emissionsTot} =\emissionsVeh + \emissionsSub.
\end{equation*}
\paragraph*{\gls{abk:mdpi} definition}
Formally:~$d_\mathrm{Mob_1}\colon \F{\mathcal{Q}}\tickar \R{\rplusbar}^3.$
The MDPI formal diagram is reported in \cref{fig:dpmob_1}.
\begin{lemma}
\label{lem:mob_1_welldef}
$d_\mathrm{Mob_1}$ is a well-defined \gls{abk:mdpi}.
\end{lemma}

\subsubsection{The Mobility \gls{abk:mdpi} (Version 2)}
\label{sec:mcdp_2}
As in \cref{sec:mcdp_1}, the functionality provided by the \gls{abk:mdpi} is the set of travel requests. 
To provide the mobility service, three resources are required. First, on the customers' side, we require an average travel time, defined in~\eqref{eq:traveltime_2}.
Second, on the side of the central authority, the resource is the total transportation cost of the intermodal mobility system. 
To the cost defined in \cref{sec:mcdp_2}, we add the cost related to \gls{abk:mm}. Assuming an average \gls{abk:mmveh}'s life of~$\lifeSco$ we get
\begin{equation*}
\R{\costTot}=\costVeh + \costSco + \costSub,
\end{equation*}
where~$\costSco$ is the \gls{abk:mmveh}-related cost
\begin{equation*}
\costSco=\frac{\costFixSco}{\lifeSco}\cdot \numberFleetSco + \costOpSco \cdot \distanceSco,
\end{equation*}
Third, we add the \gls{abk:mm}-related emissions to the ones computed in \cref{sec:mcdp_1}:
\begin{equation*}
\R{\emissionsTot} =\emissionsVeh + \emissionsSco + \emissionsSub.
\end{equation*}
\paragraph*{\gls{abk:mdpi} definition}
Formally:~$d_\mathrm{Mob_2}\colon \F{\mathcal{Q}}\tickar \R{\rplusbar}^3.$
The MDPI is reported in \cref{fig:dpmob_2}.

\begin{lemma}
\label{lem:mob_2_welldef}
$d_\mathrm{Mob_2}$ is a well-defined \gls{abk:mdpi}.
\end{lemma}

\subsubsection{The Mobility \gls{abk:mdpi} (Version 3)}
\label{sec:mcdp_3}
We now extend the setting presented in \cref{sec:mcdp_1} by including the structure presented in \cref{sec:iamod_codesign_3}.
Specifically, the functionality provided by the \gls{abk:mdpi} coincides with the functionalities of~$d_\mathrm{I_3}$ (i.e., includes travel requests and total revenue). Furthermore, to provide the functionalities the three resources introduced in \cref{sec:mcdp_1} are required:~$\averageTravelTime$ $\costTot$, and~$\emissionsTot$.
We introduce a feedback loop, by requiring the total revenue~$\rho$ to at least cover a fraction~$\chi$ of the total costs, i.e.,~$\F{\rho} \geq \chi \cdot \R{\costTot}$.\nomenclature[a100]{$\chi$}{Fraction of the revenue to cover costs}

\paragraph*{\gls{abk:mdpi} definition}
Formally:~$d_\mathrm{Mob_3}\colon \F{\mathcal{Q}}\tickar \R{\rplusbar}^2.$
The MDPI is reported in \cref{fig:dpmob_3}.

\begin{lemma}
\label{lem:mob_3_welldef}
$d_\mathrm{Mob_3}$ is a well-defined \gls{abk:mdpi}.
\end{lemma}

\subsubsection{Discussion}
First, we lump the \gls{abk:av}'s autonomy in its achievable velocity. \changed{We leave to future research more elaborated and realistic \gls{abk:av} models, accounting, for instance, for accidents rates~\cite{Richards2010} and for safety levels. For the latter, we plan on explicitly including the autonomy model, as in~\cite{zardiniIROS2020}.}
\bischanged{Similarly, the ability of \gls{abk:mm} to operate on specific network links is a regulatory aspect, which depends on factors not related to vehicle design.}
Second, we assume the service frequency of the subway system to scale monotonically with the number of trains. 
We inherently assume that the existing infrastructure can homogeneously accommodate the acquired train cars. To justify the assumption, we include an upper bound on the number of potentially acquirable trains in our case study design in \cref{sec:results}. 
Nonetheless, the co-design framework can also accommodate more sophisticated frequency models.
Third, we highlight that the intermodal mobility framework is only one of the many feasible ways to map total demand to travel time, costs, and emissions. 
\changed{Specifically, practitioners can easily replace the corresponding \gls{abk:mdpi} (here specified via a multi-commodity flow model and an optimization problem) with different models (e.g., MATSim~\cite{HorniAxhausen2016}), as long as the light condition of monotonicity of the \gls{abk:mdpi} is preserved.
In this sense, the framework is user-friendly, allowing users to plug in different models and analyze the results.}
In our setting, we conjecture customers and vehicles routes to be centrally controlled by the central authority in a socially-optimal fashion.
\bischanged{Here, investment costs have to be covered by the central authority.
We leave more complex cost analysis, and the study of strategic interactions of stakeholders, to future studies.}
Fourth, we assume a homogeneous fleet of \glspl{abk:av} and \glspl{abk:mmveh}. 
Nevertheless, our model is readily extendable to capture heterogeneous fleets.
Finally, we consider a fixed travel demand, and compute the antichain of resources providing it. 
Nonetheless, our formalization can be easily extended to arbitrary demand models preserving the monotonicity of the \gls{abk:cdpi} to account, for instance, for elastic effects~\cite{gartner1980optimal,gartner1980optimalbis}.
\section{Design of Experiments and Results}\label{sec:design of experiments and results}
\label{sec:results}
\changed{In this section, we showcase the co-design framework presented in \cref{sec:codesignav} on the case of Washington D.C., USA, leveraging real mobility data.} We detail our experimental design in \cref{sec:design of experiments} and present numerical results in \cref{subsec:results}.

\subsection{Design of Experiments}
\label{sec:design of experiments}
\bischanged{Our example study is based on the urban area of Washington D.C., USA.} 
The city road network and its features are imported from OpenStreetMap~\cite{HaklayWeber2008}, whilst the public transit network together with its schedules are extracted from GTFS~\cite{GTFS2019}. 
Original demand data is obtained by merging origin-destination pairs of the morning peak of \formatdate{1}{5}{2017}, provided by taxi companies~\cite{ODDC2017} and the Washington Metropolitan Area Transit Authority (WMATA)~\cite{PIM2012}. 
On the public transportation side, we focus our studies on the MetroRail system and its design. 
To take account of the recently increased presence of ride-hailing companies, the taxi demand rate is scaled by a factor of 5~\cite{Siddiqui2018b}.
The complete demand dataset includes \SI{16430}{} distinct origin-destination pairs, describing travel requests. 
To account for congestion effects, the nominal road capacity is computed as in~\cite{DoA1977} and an average baseline usage of \SI{93}{\percent} is assumed, in line with~\cite{DixonIrshadEtAl2018}.
We assume an \gls{abk:av} fleet composed of battery electric BEV-250 mile \glspl{abk:av}~\cite{PavlenkoSlowikEtAl2019}.
We summarize the main parameters characterizing our case studies together with their bibliographic sources in \cref{tab:params}.
In the remainder of this section, we solve the co-design problem presented in~\cref{sec:codesignav}\footnote{The solution techniques for this kind of optimization problems and their complexity are described in~\cite[Proposition 5]{Censi2015}, in the appendix, and in our talk at \url{https://bit.ly/3ellO6f}.
We are writing books on the subject, and teaching classes; see \url{https://applied-compositional-thinking.engineering}.}.
\changed{The diagrams we reported represent the ``skeleton'' of the design hierarchy.
In order to fill the blocks, one needs feasibility relations, which have been described in previous sections. 
Note that the proposed approach is extremely flexible, since it allows one to specify feasibility relations via catalogues (e.g., for vehicle models), formulas (e.g., for the cost structures), and simulation/optimization problems (e.g., for the intermodal mobility system).}
Once one identifies the \glspl{abk:mdpi}, one can directly use the PyMCDP solver~\cite{Censi2019}.
\changed{The solver provides the full set of optimal solutions. If it converges to an empty set, the solution corresponds to a certificate of infeasibility}.
Beside our basic setting (S1), we evaluate the sensitivity of the design strategies to different models of automation costs of \glspl{abk:av} (S2--S4) assess the impact of emerging \gls{abk:mm} solutions, showing how one can easily include new modes of transportation in the framework (S5), and investigate pricing strategies in (S6). We summarize the considered mobility solutions and their complementarity in Table~\ref{tab:comparisonsolutions}.

\begin{table*}[t]
	\begin{center}
	    \setlength{\tabcolsep}{4.5pt}
		\begin{scriptsize}
		\begin{tabular}{lllccccccclc}
			\toprule
			\multicolumn{2}{l}{\textbf{Parameter}} & \textbf{Variable}  & \multicolumn{7}{c}{\textbf{Value}} &\textbf{Units}& \textbf{Source}\\
			\midrule
			Road usage & & $u_{ij}$ &\multicolumn{7}{c}{\SI{93}{}} &\si{\percent}& ~\cite{DixonIrshadEtAl2018}\\
			\midrule
			&&& \textbf{S1} & \textbf{S2 (2022)} &\textbf{S2 (2025)} & \textbf{S3}& \textbf{S4} & \textbf{S5 (2022)} & \textbf{S5 (2025)}, \textbf{S6}\\ \cline{4-10}  \\[-1.0em]
			\multicolumn{2}{l}{\glspl{abk:av} operational cost} & $\costOpVeh$ &\SI{0.084}{} & \SI{0.084}{}& \SI{0.062}{} & \SI{0.084}{} & \SI{0.50}{} & \SI{0.084}{} &\SI{0.062}{} & \si{\usd\per\mile } & ~\cite{PavlenkoSlowikEtAl2019, BoeschBeckerEtAl2018}\\
			\multicolumn{2}{l}{Vehicle cost} & $\costVeh$ & \SI{32}{} & \SI{32}{} & \SI{26}{} & \SI{32}{} & \SI{32} & \SI{32} & \SI{26} &\si{k\usd\per\car}& ~\cite{PavlenkoSlowikEtAl2019}\\
			\multirow{7}{*}{\gls{abk:av} automation cost}& \SI{20}{\mph} & \multirow{7}{*}{$C_\mathrm{V,a}$} & \SI{15} & \SI{20}{} & \SI{3.7}{} & \SI{500}{} & \SI{0}{} & \SI{20}{} & \SI{3.7}{} &\si{k\usd\per\car}& ~\cite{BoeschBeckerEtAl2018,FagnantKockelman2015,BauerGreenblattEtAl2018,Litman2019,Wadud2017}\\
			 & \SI{25}{\mph} &  & \SI{15}{} & \SI{30}{} & \SI{4.4}{} & \SI{500}{} & \SI{0}{} & \SI{300}{} & \SI{4.4}{} & \si{k\usd\per\car}&~\cite{BoeschBeckerEtAl2018,FagnantKockelman2015,BauerGreenblattEtAl2018,Litman2019,Wadud2017}\\
			 & \SI{30}{\mph} &  & \SI{15}{} & \SI{55}{} & \SI{6.2}{} & \SI{500}{} & \SI{0}{} & \SI{55}{} & \SI{6.2}{} & \si{k\usd\per\car}& ~\cite{BoeschBeckerEtAl2018,FagnantKockelman2015,BauerGreenblattEtAl2018,Litman2019,Wadud2017}\\
			 & \SI{35}{\mph} &  &\SI{15}{} & \SI{90}{} & \SI{8.7}{} & \SI{500}{} & \SI{0}{} & \SI{90}{} & \SI{8.7}{} & \si{k\usd\per\car}& ~\cite{BoeschBeckerEtAl2018,FagnantKockelman2015,BauerGreenblattEtAl2018,Litman2019,Wadud2017}\\
			 & \SI{40}{\mph} &  & \SI{15}{} & \SI{115}{} & \SI{9.8}{} & \SI{500}{} & \SI{0}{} & \SI{115}{} & \SI{9.8}{} & \si{k\usd\per\car}&~\cite{BoeschBeckerEtAl2018,FagnantKockelman2015,BauerGreenblattEtAl2018,Litman2019,Wadud2017}\\
			 & \SI{45}{\mph} &  & \SI{15}{} & \SI{130}{} & \SI{12}{} & \SI{500}{} & \SI{0}{} & \SI{130}{} & \SI{12} & \si{k\usd\per\car}&~\cite{BoeschBeckerEtAl2018,FagnantKockelman2015,BauerGreenblattEtAl2018,Litman2019,Wadud2017}\\
			 & \SI{50}{\mph} &  & \SI{15}{} & \SI{150}{} & \SI{13}{} & \SI{500}{} & \SI{0}{} & \SI{150}{} & \SI{13} & \si{k\usd\per\car}& ~\cite{BoeschBeckerEtAl2018,FagnantKockelman2015,BauerGreenblattEtAl2018,Litman2019,Wadud2017}\\
			\multicolumn{2}{l}{\gls{abk:av} life} &$\lifeVeh$ & \SI{5}{} & \SI{5}{}& \SI{5}& \SI{5}{} & \SI{5}{} & \SI{5}{} & \SI{5}{} & \si{\year} &~\cite{PavlenkoSlowikEtAl2019}\\
			\multicolumn{2}{l}{CO$_2$ per Joule}& $\gamma$ & \SI{0.14}{} & \SI{0.14}{} & \SI{0.14}{} & \SI{0.14}{} & \SI{0.14}{} & \SI{0.14}{} & \SI{0.14}{} &\si{\gram\per\kilo\joule}& ~\cite{Watttime2018}\\
			\multicolumn{2}{l}{Time~$\GraphPedestrian$ to~$\GraphRoadVeh$} & $\timePedestrianRoadVeh$ & \SI{300}{} & \SI{300}{} & \SI{300}{} & \SI{300}{} & \SI{300}{} & \SI{300}{} & \SI{300}{} &\si{\second}&-\\
			\multicolumn{2}{l}{Time~$\GraphRoadVeh$ to~$\GraphPedestrian$} & $\timeRoadVehPedestrian$ & \SI{60}{} & \SI{60}{} & \SI{60}{} & \SI{60}{} & \SI{60}{} & \SI{60}{} & \SI{60}{} & \si{\second}&-\\
			\multicolumn{2}{l}{Speed limit fraction}&$\beta$&$1/1.3$&$1/1.3$&$1/1.3$&$1/1.3$&$1/1.3$&$1/1.3$&$1/1.3$&\si{\nounit}&~\cite{Dahl2018}\\
			\midrule
			&&& \textbf{\gls{abk:es}}& & \textbf{\gls{abk:sb}}& &  \textbf{\gls{abk:moped}}&  & \textbf{\gls{abk:fcm}}\\ \cline{4-10}  \\[-1.0em]
			\multicolumn{2}{l}{\gls{abk:mmveh} operational cost}&$\costOpSco$ & \SI{0.79}{} & & \SI{1.58}{} & & \SI{2.05}{} & & \SI{1.20}{} &\si{\usd\per\mile}&\cite{SchellongEtAl2019,Chao2009,DistrictColumbia2015}\\
			\multicolumn{2}{l}{\gls{abk:mmveh} cost}&$\costFixSco$ & \SI{550}{} & & \SI{8860}{} & & \SI{1000}{}& & \SI{3000}{} &\si{\usd\per\mmveh}&\cite{Ark2019,DistrictColumbia2015,Chao2009}\\
			\multicolumn{2}{l}{\gls{abk:mmveh} achievable speed}&$\arcSpeedSco$ & \SI{15}{} & & \SI{10}{} & & \SI{15}{} & & \SI{15}{} &\si{\mph}&-\\
			\multicolumn{2}{l}{\gls{abk:mmveh} life}&$\lifeSco$ & \SI{0.085}{} & & \SI{7.0}{} & & \SI{10.0}{} & & \SI{10.0}{} &\si{\year}&\cite{Ark2019,DistrictColumbia2015,Chao2009}\\
			\multicolumn{2}{l}{\gls{abk:mmveh} emissions}&$\emissionsSco$ & \SI{0.101}{} & & \SI{0.033}{} & & \SI{0.158}{} & & \SI{0.033}{} &\si{\kilogram\per\mile}&\cite{Hollingsworth2019,Kou2020,Chao2009,Van2014}\\
			\multicolumn{2}{l}{Time from $\GraphPedestrian$ to $\GraphRoadSco$ } & $\timePedestrianRoadSco$ & \SI{60}{}  & & \SI{60}{} & & \SI{60}{} & & \SI{60}{} &\si{\second}&-\\
			\multicolumn{2}{l}{Time from $\GraphRoadSco$ to $\GraphPedestrian$} & $\timeRoadScoPedestrian$ & \SI{60}{} & & \SI{60}{} & & \SI{60}{} & & \SI{60}{} &\si{\second}&-\\
			\midrule
			\multirow{3}{*}{Subway operational cost} & \SI{100}{\percent}& \multirow{3}{*}{$\costOpTrain$} & \multicolumn{7}{c}{\SI{148000000}{}} &\si{\usd\per\year}&~\cite{WMATA2017}\\
			&\SI{150}{\percent}&  & \multicolumn{7}{c}{\SI{222000000}{}} &\si{\usd\per\year}& ~\cite{WMATA2017}\\
		    &\SI{200}{\percent}&  & \multicolumn{7}{c}{\SI{295000000}{}} &\si{\usd\per\year}& ~\cite{WMATA2017}\\			
			\multicolumn{2}{l}{Subway fixed cost} & $\costFixTrain$ &  \multicolumn{7}{c}{\SI{14500000}{}}  &\si{\usd\per\train}& ~\cite{Aratani2015}\\
			\multicolumn{2}{l}{Train life}& $\lifeTrain$ & \multicolumn{7}{c}{\SI{30}{}} &{\si{\year}}& ~\cite{Aratani2015}\\ 
			\multicolumn{2}{l}{Subway emissions per train}& $\emissionsSub$ &\multicolumn{7}{c}{\SI{140000}{}} &\si{\kilogram\per\year}&~\cite{WMATA2018} \\
			\multicolumn{2}{l}{Train fleet baseline} & $\numberFleetTrainBaseline$ &\multicolumn{7}{c}{\SI{112}{}}& \si{\train}& ~\cite{Aratani2015}\\
			\multicolumn{2}{l}{Subway service frequency}& $\varphi_{j,\mathrm{baseline}}$ & \multicolumn{7}{c}{$1/6$}& \si{\per\minute}& -\\
			\multicolumn{2}{l}{Time~$\GraphPedestrian$ to~$\mathcal{G}_\mathrm{P}$} & $\timePedestrianSubway$ & \multicolumn{7}{c}{\SI{60}{}}& \si{\second}&-\\
		\bottomrule
	\end{tabular}
	\caption{Parameters, variables, numbers, and units for the case studies.}
	\label{tab:params}
\end{scriptsize}
\end{center}
\end{table*}

\begin{table*}[t]
    \begin{center}
    \begin{tabular}{llllll}
    \toprule
    &Mobility Type&Emissions&Cost&Speed&Reliability\\
    \midrule
    Taxi&Point-to-point&High&High operational cost, medium fixed cost&High&Up to availability and congestion\\
    \gls{abk:av}&Point-to-point&High&Low operational cost, high fixed cost&High&Up to availability and congestion\\
    \gls{abk:mmveh}&Point-to-point&Medium&Medium operational cost, low fixed cost&Low/Medium&Up to availability\\
    Walking&Point-to-point&No emissions&Free&Low&High\\
    Subway&Fixed hubs and routes&Low&Low&Medium&High \\
    \bottomrule 
    \end{tabular}
    \caption{Comparison of the considered mobility solutions.}
    \label{tab:comparisonsolutions}
    \end{center}
\end{table*}

\begin{LaTeXdescription}
		\item[S1 - Basic setting:] We consider the co-design of the mobility system by means of \gls{abk:amod} and public transportation systems (\cref{sec:mcdp_1}), and do not include \gls{abk:mm} solutions (cf. S5). Specifically, we 
		co-design the system by means of the \gls{abk:av} fleet size, achievable free-flow speed (see Lemma~\ref{lem:red_road_monotone}), and subway service frequency (see Lemma~\ref{lem:red_pub_monotone}): The municipality is allowed to (i) deploy an \gls{abk:av} fleet of size $\numberFleetVeh\in \{\SI{0}{},\SI{500}{},\SI{1000}{},\ldots,\SI{5000}{}\}$ vehicles, (ii) choose the single \gls{abk:av} achievable speed (determining the serviced mobility network)~$\achievableSpeedVeh\in \{\SI{20}{\mph},\SI{25}{\mph},\ldots,\SI{50}{\mph}\}$, and (iii) increase the subway service frequency~$\freqTrain$ by a factor of \SI{0}{\percent}, \SI{50}{\percent}, or \SI{100}{\percent}. 
		In line with recent literature~\cite{BoeschBeckerEtAl2018,FagnantKockelman2015,BauerGreenblattEtAl2018,Wadud2017, Litman2019,becker2020impact}, we assume an average achievable-velocity-independent cost of automation.
		
		\item[S2 - Speed-dependent automation costs:] To relax the potentially unrealistic assumption of a velocity-independent automation cost, we consider a performance-dependent cost structure, detailed in \cref{tab:params}. The large variance in sensing technologies available on the market and their performances suggests that \gls{abk:av} costs are, in fact, performance-dependent~\cite{GawronKeoleianEtAl2018, zardiniIROS2020}.
		Indeed, the technology currently required to safely operate an autonomous vehicle at \SI{50}{\mph} is substantially more sophisticated, and therefore more expensive, than the one needed at \SI{20}{\mph}. Furthermore, the frenetic evolution of automation techniques will inevitably reduce automation costs: Experts forecast a massive automation cost reduction (up to \SI{90}{\percent}) in the next decade, principally due to mass-production of \glspl{abk:av} sensing technology~\cite{WCP2018,Lienert2019}. 
		Therefore, we perform our studies with current (2022) automation costs as well as with their projections for the upcoming years (2025)~\cite{Lienert2019,PavlenkoSlowikEtAl2019,becker2020impact}.

		\item[S3 - High automation costs:] \changed{We assess the impact of high automation costs. In particular, we assume a performance-independent automation cost of \SI{0.5}{\million\usd\per\car}, capturing the extremely high research and development costs that \glspl{abk:av} companies are facing today~\cite{Korosec2019}, as well as insurance costs and infrastructural investments.
		The latter, often referred to as ``autonomy-enabling infrastructure'', would allow high driving speeds, and could consist of dedicated roads, equipped with sensors and cloud computing capabilities, enhancing the performance of \glspl{abk:av}.}

		\item[S4 - \acrshort{abk:mod} setting:] We analyze the current \gls{abk:mod} case. The cost structure of \gls{abk:mod} systems is characterized by lower vehicle costs (due to lack of automation) and higher operation costs, mainly due to drivers' salaries.
		
		\item[S5 - Impact of new transportation modes:] We show the modularity of our framework by evaluating the impact of \gls{abk:mm} solutions on urban mobility (\cref{sec:mcdp_2}). 
		We consider \glspl{abk:es} (e.g., Lime in DC), \glspl{abk:sb} (e.g., Capital Bikeshare in DC), \glspl{abk:moped} (e.g., Revel in DC), and \glspl{abk:fcm}. 
		In addition to the design parameters introduced in the basic setting, we design the specific \gls{abk:mm} solution~$M\in \{$\gls{abk:es},\gls{abk:sb},\gls{abk:moped},\gls{abk:fcm}$\}$ and the \gls{abk:mm} fleet size $\numberFleetSco \in \{0,500,\SI{1000},\ldots,\SI{5000}\}$ vehicles (see Lemma~\ref{lem:red_veh_micro_monotone}). We study the joint deployment of \gls{abk:mm} solutions and \glspl{abk:av}, and therefore consider the extended settings of 2022 and 2025. 
		\item[S6 - Pricing:] We show another extension of our framework to capture pricing strategies and infrastructure-contributing revenues (\cref{sec:mcdp_3}) in the 2022 setting. 
		We consider \gls{abk:amod} service providers that choose from an exemplary set of prices~$\{0.8, 1.2, 1.6, 2.4, 3.2\}$ (expressed in~USD/mile) and public transit authorities choosing fare prices from the set~$\{1.0, 2.0,4.0,6.0\}$ (expressed in USD per ride).
		\changed{Furthermore, we consider a municipality willing to cover~50\% (just a particular choice) of the investment cost through the revenues of mobility services. (i.e., the revenue gained from travelers paying for the trips should at least be enough to cover 50\% of the investment costs).}
\end{LaTeXdescription}

\subsection{Results}
\label{subsec:results}
\subsubsection{Basic setting}
\label{subsec:basic setting}
\begin{figure}[tbh]
    \begin{center}
    \begin{subfigure}[h]{\columnwidth}
    \includegraphics[width=\columnwidth]{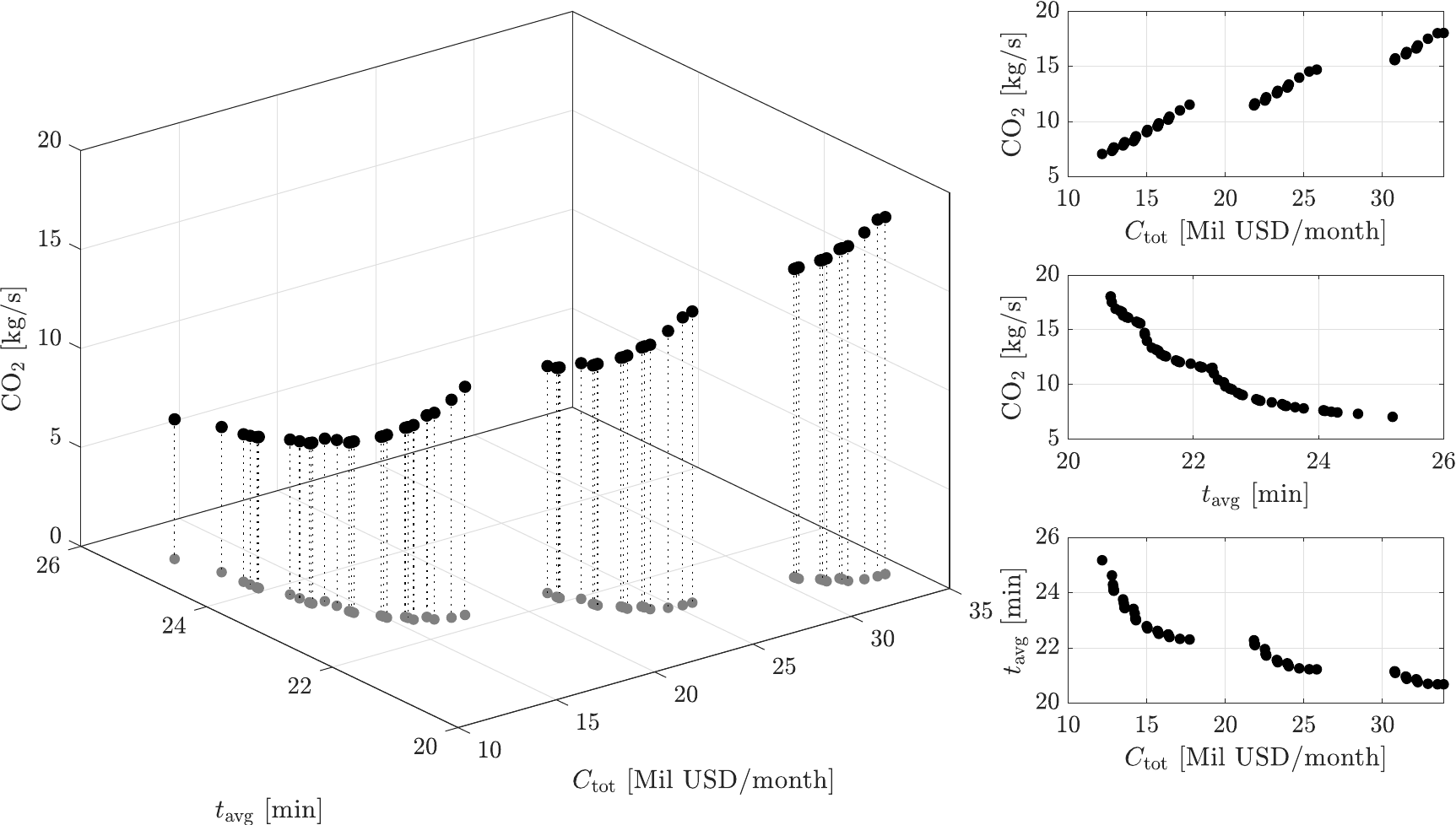}
    \caption{Left: Three-dimensional representation of antichain elements and their projection in the cost-time space. Right: Two-dimensional projections.}
    \label{fig:resultsart3D}
    \end{subfigure}
    ~
    \begin{subfigure}[h]{\columnwidth}
	\includegraphics[width=\columnwidth]{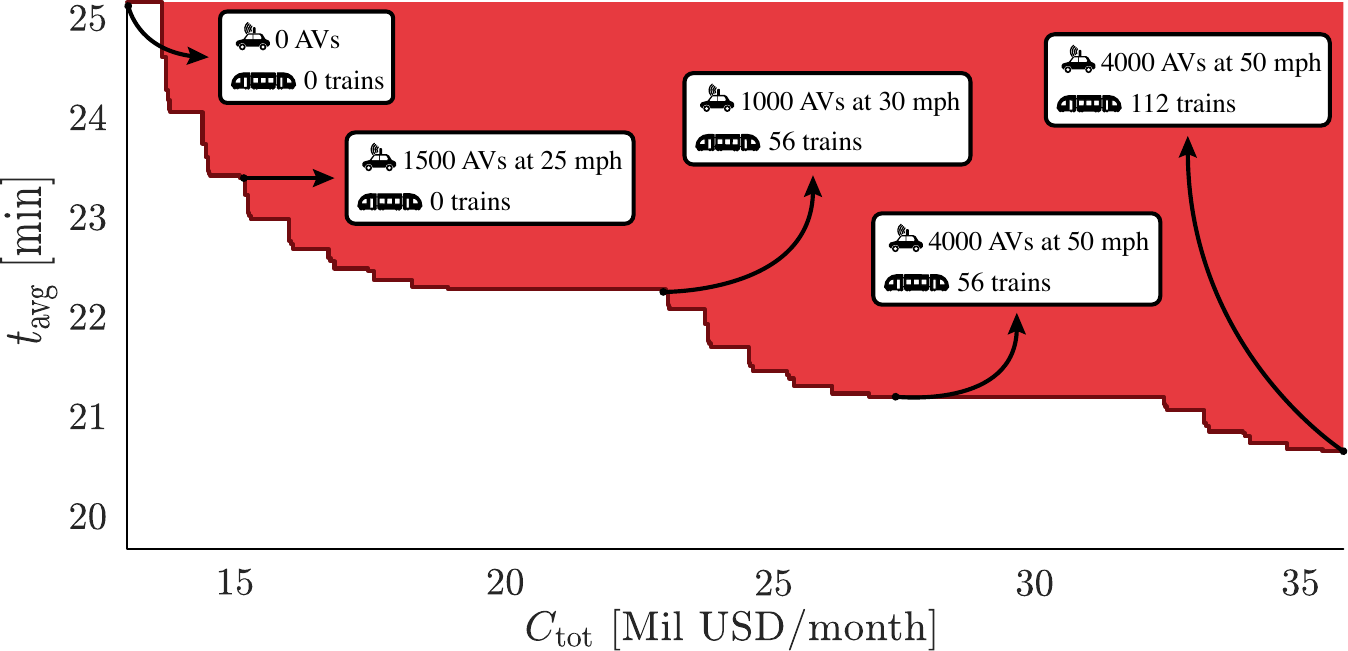}
    \caption{Results for constant automation costs. We report the two-dimensional representation of the antichain elements: The pareto front is represented in dark red, and the upper set is the area above.
    We also report selected implementations corresponding to the highlighted antichain elements, in this case quantified in terms of achievable vehicle speed, \glspl{abk:av} fleet size, and train fleet size\protect\footnotemark.}
    \label{fig:resultsart2D}
    \end{subfigure}
    \caption{Solution of the \gls{abk:cdpi}: Basic setting.}
    \label{fig:resultsart}
    \end{center}
\end{figure}

\cref{fig:resultsart3D} reports the solution of the co-design problem through the antichain consisting of the total $\text{CO}_2$ emissions, average travel time, and total transportation cost.
The design solutions are \emph{rational} (and not comparable), since there exists no instance which simultaneously yields lower emissions, average travel time, and cost.
\footnotetext{The description in this caption is valid for all the following figures.}
In the interest of clarity, we prefer a two-dimensional antichain representation, where emissions are included in the costs via a conversion factor of \SI{40}{\usd\per\kilogram}~\cite{HowardSylvan2015}. Note that this transformation preserves the monotonicity of the \gls{abk:cdpi} and therefore integrates in our framework.
The two-dimensional antichain and the corresponding central authority's decisions are reported in~\cref{fig:resultsart2D}.
In general, as the municipality budget increases, the average travel time per trip required to satisfy the given demand decreases, reaching a minimum of about \SI{20.7}{\minute}, with a monthly public expense of around \SI{36}{\million\usd\per\month}. This configuration corresponds to a fleet of \SI{4000}{} \glspl{abk:av} able to drive at \SI{50}{\mph}, and to the doubling of the current MetroRail train fleet. Furthermore, the smallest rational investment of \SI{13}{\million\usd\per\month} leads to a \SI{22}{\percent} higher average travel time, corresponding to the current situation, i.e., to a non-existent \glspl{abk:av} fleet, and an unchanged subway infrastructure. Notably, an expense of \SI{18}{\million\usd\per\month} (\SI{50}{\percent} lower than the highest rational investment) only increases the minimal required travel time by \SI{8}{\percent}, requiring a fleet of \SI{3000}{} \glspl{abk:av} able to drive at \SI{45}{\mph} and no acquisition of trains. Conversely, an expense of \SI{15}{\million\usd\per\month} (just \SI{2}{\million\usd\per\month} higher than the minimal rational investment) provides a \SI{2}{\minute} shorter travel time. Finally, it is rational to improve the subway system starting from a budget of \SI{23}{\million\usd\per\month}, leading to a travel improvement of just \SI{8}{\percent}. This trend can be explained with the high train acquisition cost and increased operation costs, related to the reinforcement of the subway system. This phenomenon is expected to be even more marked for other cities, considering the moderate operation costs of the MetroRail subway system, due to its automation and related benefits~\cite{WangZhangEtAl2016}.

\subsubsection{Speed-dependent automation costs}
\subsubsection*{2022}
\begin{figure}[tb]
    \begin{center}
    \begin{subfigure}[h]{\columnwidth}
    \includegraphics[width=\columnwidth]{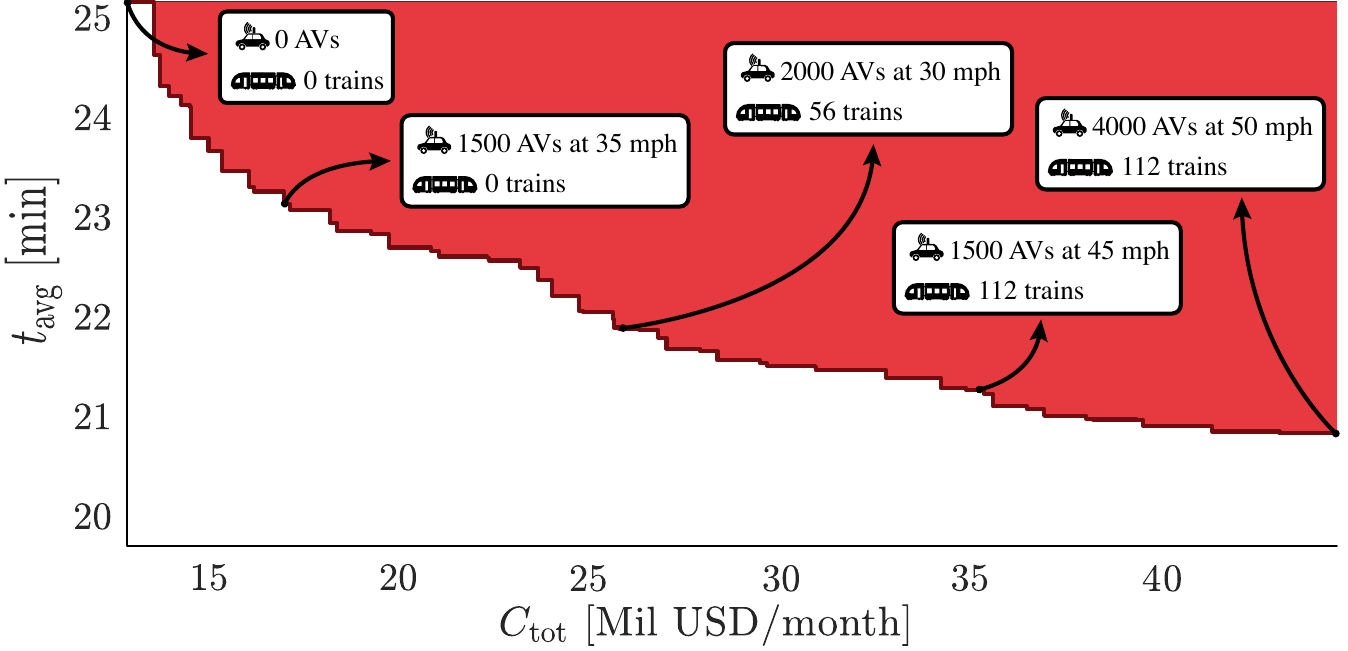}
    \caption{Speed-dependent automation costs in 2022.}
    \label{fig:results19}
     \end{subfigure}
     ~
     \begin{subfigure}[h]{\columnwidth}
    \includegraphics[width=\columnwidth]{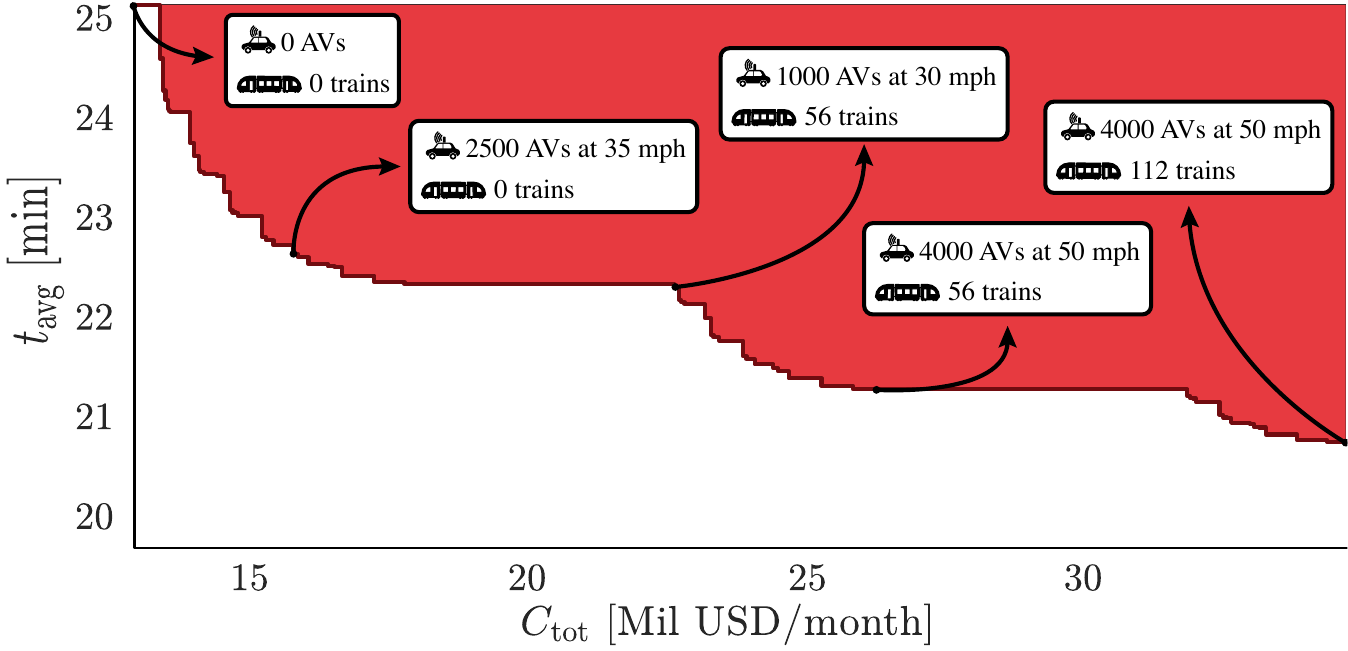}
    \caption{Speed-dependent automation costs in 2025.}
    \label{fig:results25}
     \end{subfigure}
    \caption{Results for the speed-dependent automation costs.}
    \label{fig:results1925}
    \end{center}
\end{figure}

We report the results in~\cref{fig:results19}.
A comparison with our basic setting (cf. \cref{fig:resultsart})  confirms the trends concerning public expense. Indeed, a public expense of \SI{26}{\million\usd\per\month} (\SI{43}{\percent} lower than the highest rational expense) only increases the average travel time by \SI{5}{\percent}, requiring a fleet of \SI{2000}{} \glspl{abk:av} able to reach \SI{30}{\mph} and a subway reinforcement of \SI{50}{\percent}. Nevertheless, our comparison shows two substantial differences.
First, the budget required for an average travel time of \SI{13}{\minute} is \SI{25}{\percent} higher compared to S1.
Second, the higher \gls{abk:av} costs result in an average \glspl{abk:av} fleet growth of \SI{9}{\percent}, an average velocity reduction of \SI{15}{\percent}, and an average train fleet growth of \SI{14}{\percent}. The latter suggests a shift towards poorer \glspl{abk:av} performance in favor of fleets reinforcements.

\subsubsection*{2025}
The maximal rational budget is \SI{23}{\percent} lower than in the case of immediate deployment (\cref{fig:results25}).
Further, the reduction in autonomy costs incentifies the acquisition of more performant \glspl{abk:av}, increasing the average vehicle speed by \SI{14}{\percent}. Hence, \glspl{abk:av} and train fleets are \SI{10}{\percent} and \SI{13}{\percent} smaller.

\subsubsection{High automation costs analysis}
\begin{figure}[tb]
    \begin{center}
    \includegraphics[width=\columnwidth]{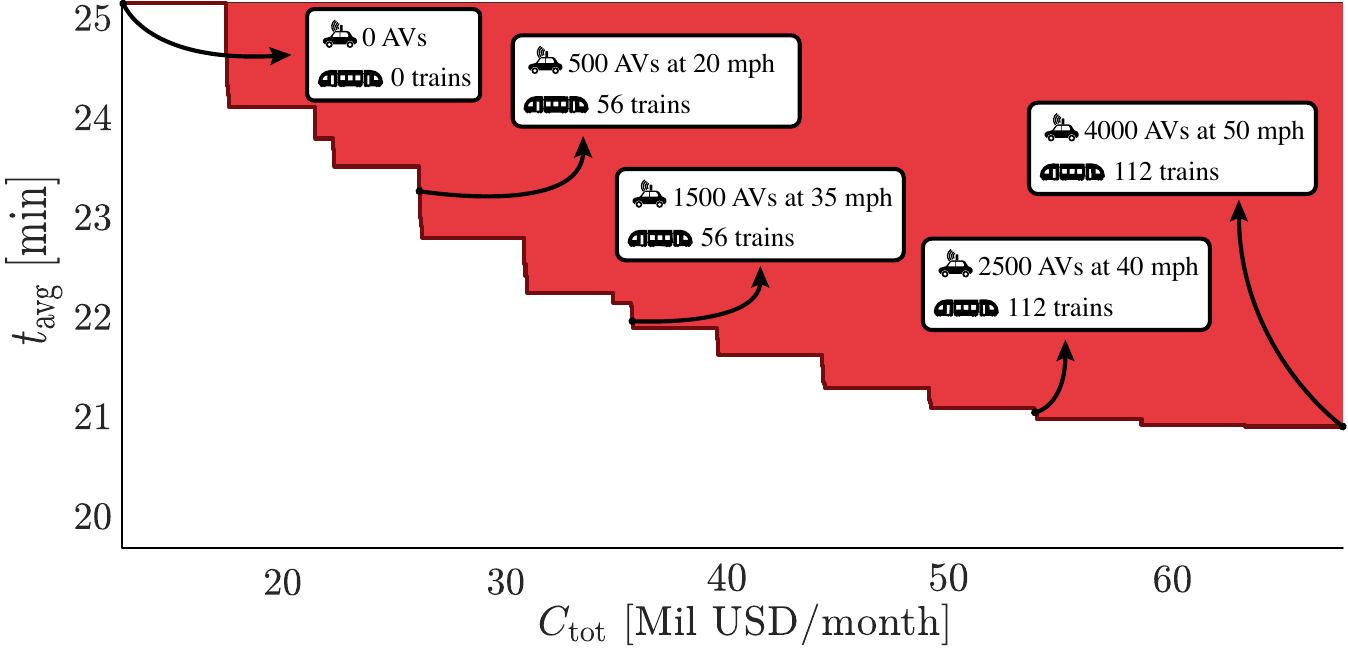}
    \end{center}
    \caption{Results for large automation costs.}
    \label{fig:resultshigh}
\end{figure}
\changed{\cref{fig:resultshigh} shows the results for high automation costs.} 
First, we observe a substantial shift towards larger train fleet sizes (\SI{65}{\percent} larger than in S1) and smaller \glspl{abk:av} fleets (\SI{55}{\percent} smaller than in S1).
Second, minimizing the average travel time entails an expense of approximately \SI{68}{\million\usd\per\month}, basically doubling the investments observed in the basic setting.

\subsubsection{\gls{abk:mod} setting}
\begin{figure}[tb]
    \begin{center}
    \includegraphics[width=\columnwidth]{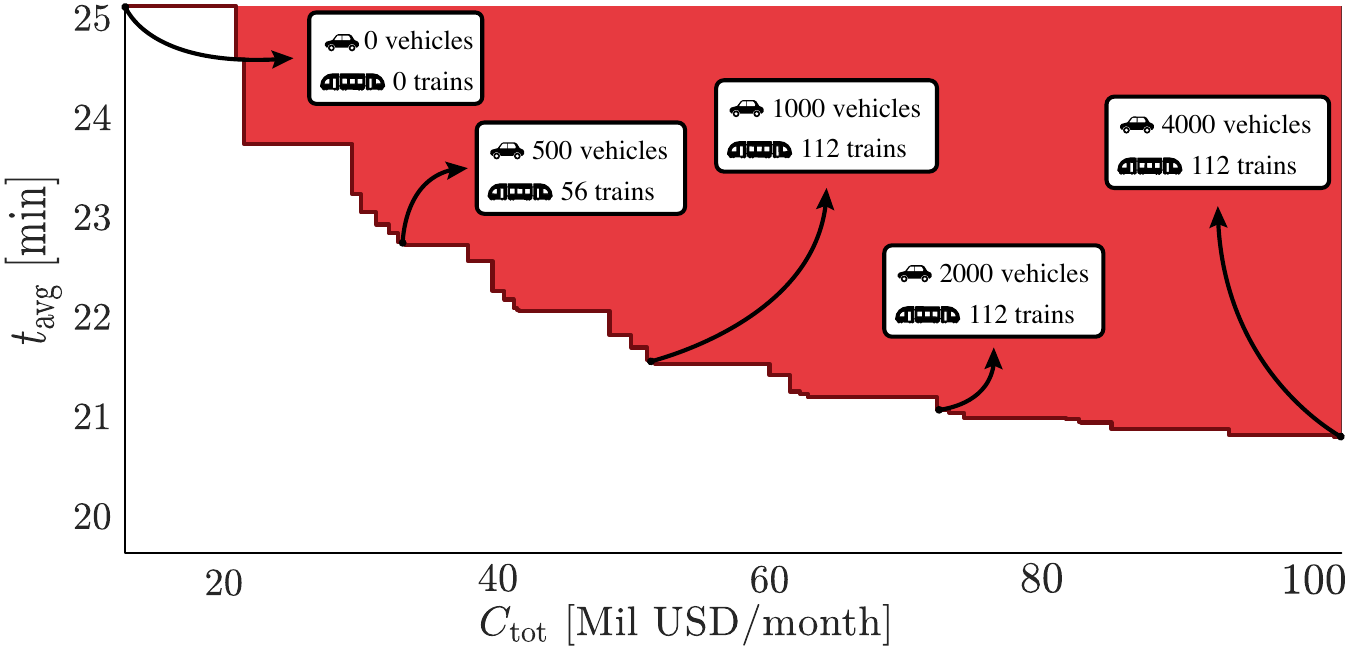}
    \caption{Results for the \gls{abk:mod} case.}
    \label{fig:resultsmod}
    \end{center}
\end{figure}
We summarize the results for the \gls{abk:mod} scenario in \cref{fig:resultsmod}. In particular, by comparing the \gls{abk:mod} case with the 2025 setting, we can notice the game-changing properties that \glspl{abk:av} introduce in the mobility ecosystem. In particular, the average train fleet size and the average vehicle fleet sizes increase by \SI{130}{\percent} and \SI{66}{\percent}, suggesting a clear transition in investments from public transit to \glspl{abk:av}, and testifies to the interest in \gls{abk:amod} systems developed in the past years.

\subsubsection{Impact of new transportation modes}
\begin{figure}[tb]
    \begin{center}
    \begin{subfigure}[b]{\columnwidth}
    \includegraphics[width=\columnwidth]{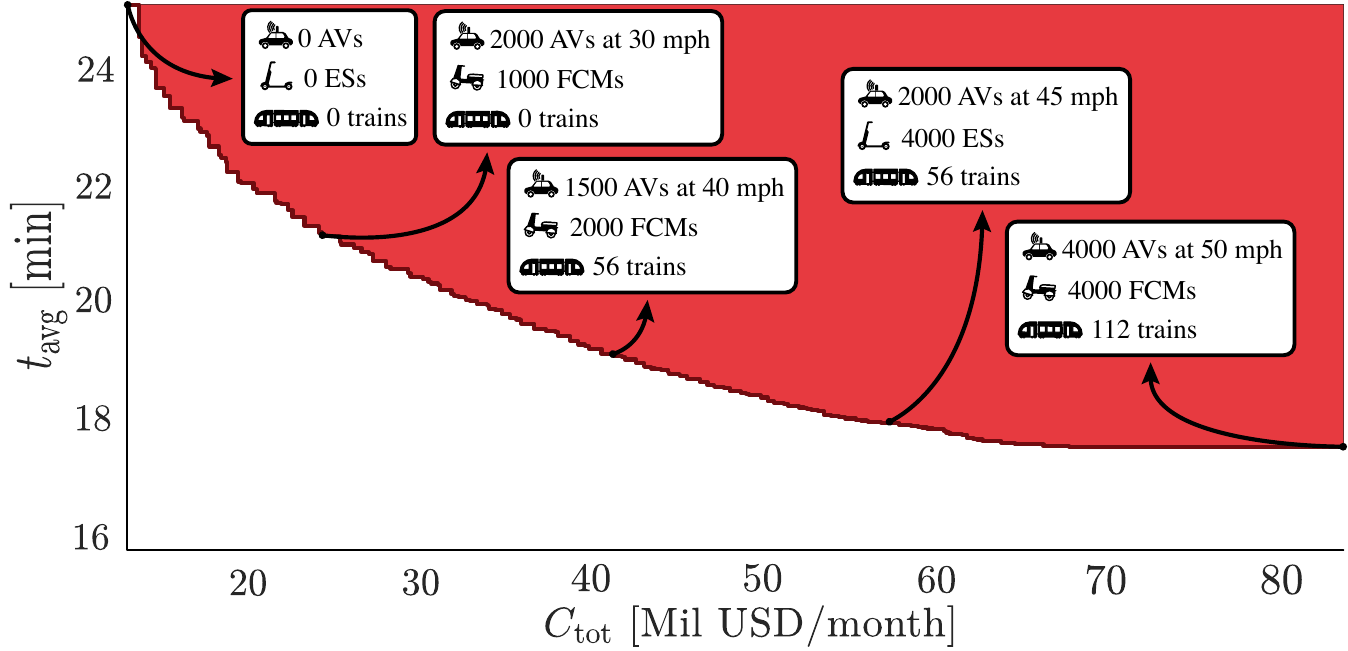}
    \caption{Impact of micromobility in 2022.}
    \label{fig:resultsmoped2020}
     \end{subfigure}
     ~
     \begin{subfigure}[b]{\columnwidth}
    \includegraphics[width=\columnwidth]{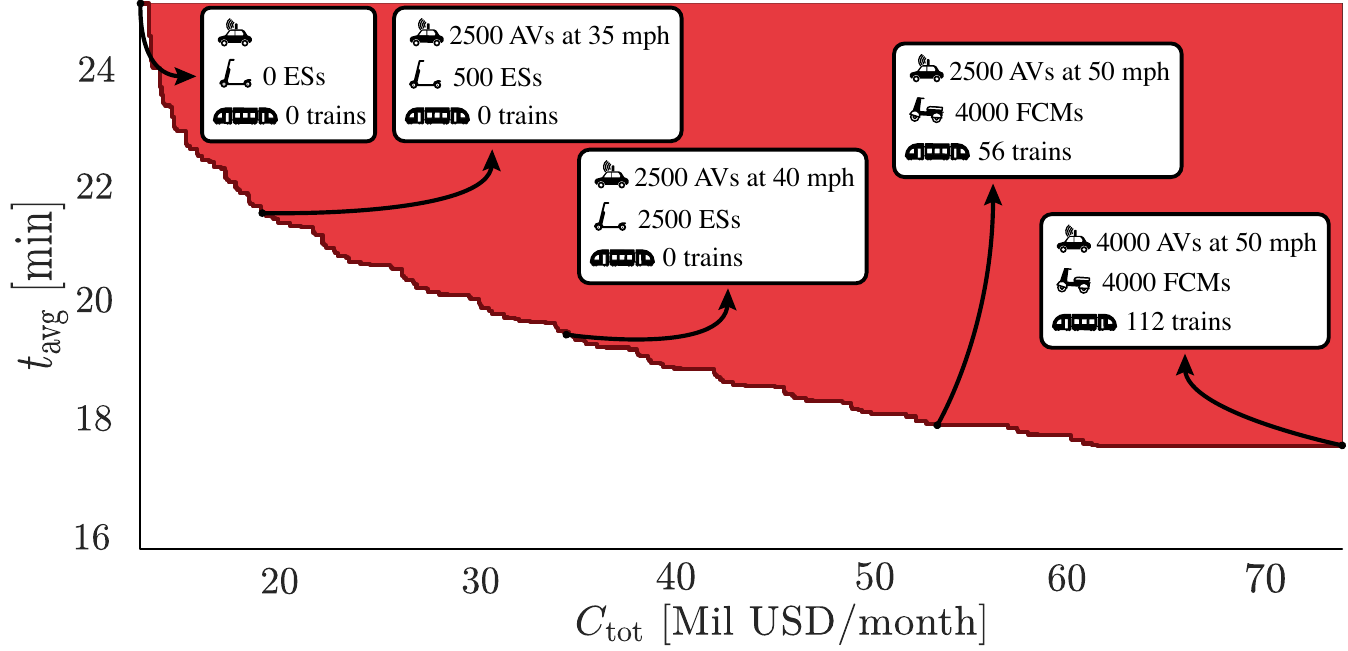}
    \caption{Impact of micromobility in 2025.}
    \label{fig:resultsmoped2025}
     \end{subfigure}
    \end{center}
    \caption{Results for the impact of micromobility.}
    \label{fig:resultsmopeds}
\end{figure}
To assess the impact of \gls{abk:mm} solutions, we compare the arising design solutions, reported in \cref{fig:resultsmopeds}, with their counterpart in S2 (cf. \cref{fig:results1925}).

\subsubsection*{2022}
\cref{fig:resultsmoped2020}, together with \cref{fig:results19}, demonstrates an overall benefit from \gls{abk:mm} solutions. For instance, the most time-efficient solution in S2 yields an average travel time of \SI{20.7}{\minute} at an expense of \SI{45}{\million\usd\per\month}. The deployment of \gls{abk:mm} solutions lowers the average travel time achievable with the same expense by \SI{10}{\percent} (\SI{18.8}{\minute}) and allows for even lower average travel times, with a time-efficient solution of \SI{17.6}{\minute} at an investment plan of \SI{84}{\million\usd\per\month}. Overall, the average \glspl{abk:av} fleet size and the average train fleet size are \SI{35}{\percent} and \SI{6}{\percent} smaller, in favor of an average \gls{abk:mm} fleet of \SI{2280}{} \glspl{abk:mmveh}.

\subsubsection*{2025}
\cref{fig:resultsmoped2025}, together with \cref{fig:results25}, shows that the benefit of \gls{abk:mm} solutions is less marked than in 2022. 
For instance, an expense of \SI{35}{\million\usd\per\month} (same as the maximal expense in~\cref{fig:results25}) results in an average travel time of \SI{19.5}{\minute}, i.e., only \SI{6}{\percent} lower than in the case without \gls{abk:mm}.
Furthermore, we observe an average \glspl{abk:av} fleet size enlargement of \SI{17}{\percent}, and an average train fleet size reduction of \SI{27}{\percent}.
Finally, the comparison with the 2022 case highlights a \glspl{abk:mmveh} fleet reduction of \SI{23}{\percent}, which suggests the comparative advantage of \glspl{abk:av} in the future. Indeed, the stronger the reduction of the cost of automation, the more investments in \glspl{abk:av} are rational.
The benefits of employing \gls{abk:mm} solutions could therefore just be temporary, and gradually vanish as the costs of automation of \glspl{abk:av} decrease.

\subsubsection{Pricing}
\begin{figure}[tb]
\begin{center}
\includegraphics[width=\columnwidth]{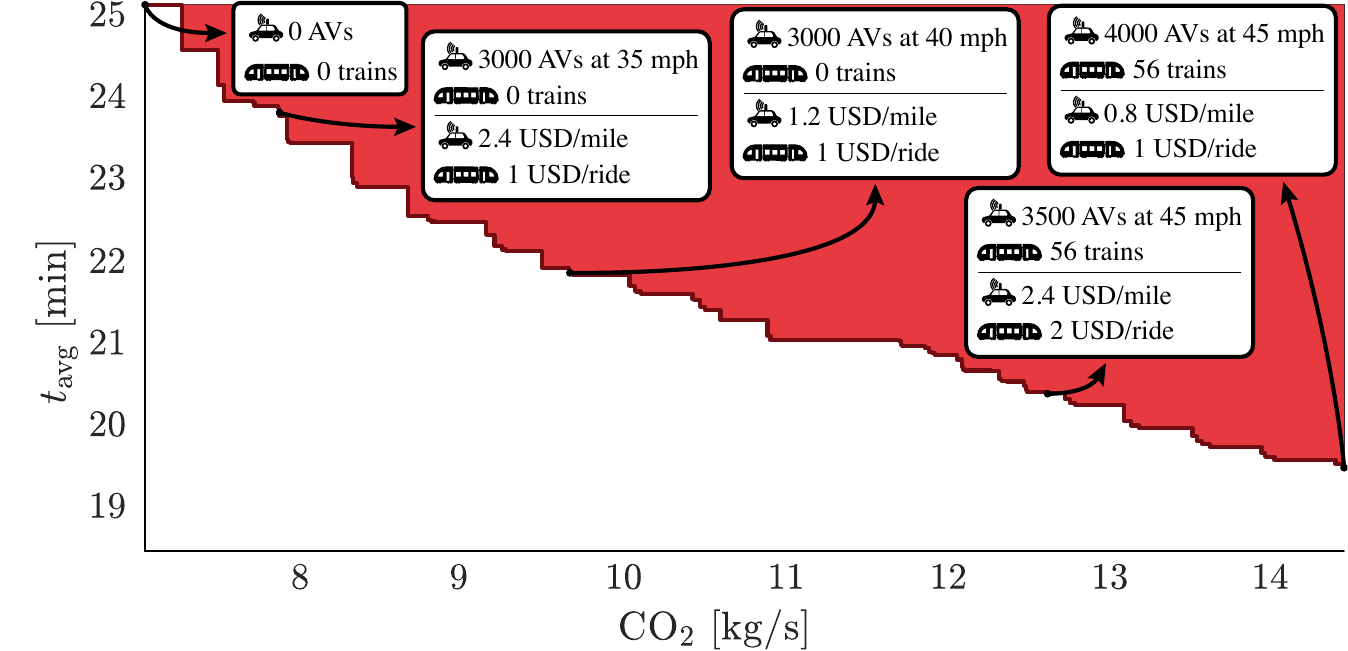}
\end{center}
\caption{Pricing and revenues case study. The Pareto front is in terms of system performance (average travel time) and produced externalities.}
\label{fig:resultspricing}
\end{figure}

We report the results in \cref{fig:resultspricing}. 
In particular, we report the Pareto front between system performance and emissions (the two resources of the considered \gls{abk:mdpi}), as well as design choices for selected Pareto-optimal solutions, now including prices for \gls{abk:amod} and public transit services.
We report three key observations.
First, the most performing solution (which satisfies the cost-contributing constraint) features the usage of \SI{4000}~\glspl{abk:av} able to drive at \SI{45}{\mph} and an increment of 50\% of the public transit fleets.
The large usage of \glspl{abk:av} is not only due to their efficiency, but also to the low price of \SI{0.8}{\usd/\mile}.
While this choice does not fully exploit the action space of the municipality (one could have larger fleets, more performant \glspl{abk:av}, and more trains), it is the last one for which the weighted costs do not exceed the revenue.
Second, we observe fewer solutions featuring an augmented train fleet, mainly because of the related onerous investments related (i.e., more \glspl{abk:av}, not necessarily very performing, can bridge the system performance gap).
Finally, comparing the emissions in \cref{fig:resultspricing} and in \cref{fig:resultsart3D} suggests that bounding the allowed mobility system costs also prevents design options which are more pollutant from being chosen.

\subsection{Discussion}
First, the presented case studies showcase the ability of our framework to extract the set of rational design strategies for a future mobility system, including \glspl{abk:av}, \glspl{abk:mmveh}, and public transit. 
\changed{This way, stakeholders such as mobility providers, transportation authorities, and policy makers can get transparent and interpretable insights on the impact of future interventions, inducing further reflection on this complex socio-technical problem.
Note that this kind of results is only one of the many factors affecting negotiations when interacting with stakeholders.}
Second, we perform a sensitivity analysis through the variation of autonomy cost structures, and show the capacity of our framework to capture various models. On the one hand, this reveals a clear transition from small fleets of fast \glspl{abk:av} (in the case of low autonomy costs) to large fleets of slow \glspl{abk:av} (in the case of high autonomy costs). On the other hand, our studies highlight that investments in the subway infrastructure are rational only when large budgets are available. Indeed, the high train acquisition and operation costs lead to a comparative advantage of \gls{abk:av}-based mobility.
Finally, our case studies suggest that the deployment of \gls{abk:mm} solutions is rational primarily on a short-term horizon: The lowering of automation costs could eventually make \glspl{abk:av} the predominant actor in the future of urban mobility.
\section{Conclusion}\label{sec:conclusion}
This paper leverages the mathematical theory of co-design to propose a co-design framework for future mobility systems. The nature of our framework offers a different viewpoint on the future mobility problem, enabling the modular and compositional interconnection of the design problems of different mobility options and their optimization, given multiple objectives. Starting from the multi-commodity flow model of an intermodal mobility system, we designed \glspl{abk:av}, \glspl{abk:mmveh}, and public transit both from a vehicle-centric and fleet-level perspective. Specifically, we studied the problem of deploying a fleet of self-driving vehicles providing on-demand mobility in cooperation with \gls{abk:mm} solutions and public transit, adapting the speed achievable by \glspl{abk:av} and \glspl{abk:mmveh}, their fleet sizes, and the service frequency of the subway lines. 
Our framework allows stakeholders involved in the mobility ecosystem, from vehicle developers all the way to mobility-as-a-service companies and central authorities, to characterize rational trajectories for technology and investment development. 
\changed{We showcased both the developer and the user views of the framework, explaining how practitioners can easily use their models within it.}
\bischanged{The proposed methodology is showcased in a case study based on data for Washington D.C., USA.} Notably, we highlighted how our problem formulation allows for a systematic analysis of incomparable objectives, such as public expense, average travel time, and emissions, providing stakeholders with analytical insights for the socio-technical design of future mobility systems. This work urges the following future research streams:
\subsubsection*{Modeling} 
First, we would like to capture heterogeneous fleets of \glspl{abk:av}, with different autonomy pipelines, propulsion systems, and passenger capacity. For instance, the modular nature of the framework allows one to easily include complex autonomy models in the design problem of the \gls{abk:av} fleet~\cite{zardiniIROS2020,zardiniECC21}.
Second, we would like to investigate variable demand models. Third, we would like to analyze the interactions between multiple stakeholders in the mobility ecosystem, characterized by conflicting interests and different action spaces. It is advantageous to formulate this as a game, and to characterize potentially arising equilibria~\cite{lanzettiEtAl2019,zardinilanzetti2021}, possibly leveraging recent results in posetal games~\cite{ZZPosetal}.
\changed{This might bring realism and effectiveness in the actionable information proposed to the mobility stakeholders.}
\bischanged{Finally, we would like to explicitly include (and not just via costs) more elements of urban design in the co-design model, to account for more realistic scenarios~\cite{maheshwari2020urban}. These include parking spaces and autonomy-enabling infrastructure.}
\subsubsection*{Algorithms}
We are interested in tailoring general co-design algorithmic frameworks to the particular case of transportation design problems, leveraging their specific structure, and characterizing their solutions. In particular, we would like to study adaptive approaches to cleverly simulate mobility systems.
\begin{appendix}
\changed{
\subsection{Nomenclature}

\vspace{-1cm}

\renewcommand{\nomname}{}
\bischanged{\printnomenclature}}
\changed{\subsection{Background on solution of co-design problems}
\label{sec:orders_background}

For the convenience of the reader, in the following we report technical results from \cite{Censi2015,censi2022}.

\subsubsection{Solution of \gls{abk:cdpi}}

We first recall concepts related to fixed points.
\begin{definition}[Least fixed point]
A \emph{least fixed point} of $f\colon\mathcal{P}\to \mathcal{P}$ is the minimum (if it exists) of the set of fixed points of $f$:
\begin{equation*}
\text{lfp}(f)=\min_{\preceq_\mathcal{P}}\{x\in \mathcal{P}\colon f(x)=x\}.
\end{equation*}
\end{definition}
A least fixed point might not exist. Monotonicity of the map $f$ and completeness of the partial orders is sufficient to ensure existence.

\begin{definition}[Completeness]
A poset is a \emph{directed complete partial order (DCPO)} if each of its directed subsets has a supremum (least of upper bounds). It is a \emph{complete partial order (CPO)} if it also has a bottom.
\end{definition}
\begin{example}
Consider~$\mathbb{R}_{\geq 0}=\{x\in \mathbb{R}\mid x\geq 0\}$, which has a bottom~$\bot=0$.
One can make~$\tup{\mathbb{R}_{\geq 0},\leq}$ a CPO by adding an artificial top element~$\top$, by defining~$\overline{\mathbb{R}}_{\geq 0}\coloneqq \mathbb{R}_{\geq 0}\cup \{\top\}$, and extending the partial order~$\leq$ such that~$a\leq \top$ for all~$a\in \mathbb{R}_{\geq 0}$.
\end{example}

\begin{lemma}[Lemma 3 in \cite{Censi2015}]
If $\mathcal{P}$ is a CPO and $f\colon \mathcal{P}\to \mathcal{P}$ is monotone, then $\text{lfp}(f)$ exists.
\end{lemma}
Assuming Scott continuity of $f$, Kleene's algorithm is a systematic procedure to find the least fixed point.

\begin{definition}[Scott continuity]
A map~$f\colon \mathcal{P}\to \mathcal{Q}$ between DCPOs is \emph{Scott continuous} if and only if, for each directed subset~$D\subseteq \mathcal{P}$, the image~$f(D)$ is directed, and~$f(\text{sup }D)=\text{sup }f(D)$.

\end{definition}

\begin{lemma}[Lemma 4 in \cite{Censi2015}]
Assume $\mathcal{P}$ is a CPO and $f\colon \mathcal{P}\to \mathcal{P}$ is Scott continuous. 
Then, the least fixed point of $f$ is the supremum of the Kleene ascent chain
\begin{equation*}
\bot \preceq f(\bot)\preceq f(f(\bot))\preceq \ldots \preceq f^{(n)}(\bot)\preceq \ldots.
\end{equation*}
\end{lemma}
Note that a sufficient condition is to assume all posets to be finite.
\begin{theorem}
The map $h$ for a \gls{abk:cdpi} has an explicit expression in terms of the maps $h_d$ of its subproblems.
\end{theorem}
We report the specific definition of these maps.
\begin{definition}[Series]
For two maps $h_1\colon \setOfFunctionalities{1}\to \mathsf{A}\setOfResources{1}$,~$h_2\colon \setOfFunctionalities{2}\to \mathsf{A}\setOfResources{2}$, if $\setOfResources{1}=\setOfFunctionalities{2}$, define
\begin{equation*}
\begin{aligned}
h_1\fatsemi h_2\colon \setOfFunctionalities{1}&\to \mathsf{A}\setOfResources{2}\\
\F{f_1}&\mapsto \Min_{\preceq_{\setOfResources{2}}}\bigcup_{s\in h_1(\F{f})}h_2(s).
\end{aligned}
\end{equation*}
\end{definition}
\begin{definition}[Parallel]
For two maps $h_1\colon \setOfFunctionalities{1}\to \mathsf{A}\setOfResources{1}$,~$h_2\colon \setOfFunctionalities{2}\to \mathsf{A}\setOfResources{2}$, define
\begin{equation*}
\begin{aligned}
h_1\otimes h_2\colon \setOfFunctionalities{1}\times \setOfFunctionalities{2}&\to \mathsf{A}(\setOfResources{1}\times \setOfResources{2})\\
\tup{\F{f_1},\F{f_2}}&\mapsto h_1(\F{f_1})\times h_2(\F{f_2}).
\end{aligned}
\end{equation*}
\end{definition}
\begin{definition}[Loop]
For  $h\colon \setOfFunctionalities{}\times \setOfResources{}\to \mathsf{A}\setOfResources{}$ define
\begin{equation*}
\begin{aligned}
h^\dagger \colon \setOfFunctionalities{}&\to \mathsf{A} \setOfResources{}\\
\F{f}&\mapsto \text{lfp}(\Psi_{\F{f}}^h),
\end{aligned}
\end{equation*}
where
\begin{equation*}
\begin{aligned}
\Psi_{\F{f}}^{h}\colon \mathsf{A}\setOfResources{}&\to \mathsf{A}\setOfResources{}\\
\R{R}&\mapsto \Min_{\preceq_{\setOfResources{}}}\bigcup_{\R{r}\in \R{R}}h(\F{f},\R{r})\cap \uparrow \{\R{r}\},
\end{aligned}
\end{equation*}
where $\uparrow$ represents the upper closure operator.
\end{definition}
\begin{definition}[Coproduct]
For $h_1,h_2\colon \setOfFunctionalities{\to \mathsf{A}\setOfResources{}}$, define
\begin{equation*}
\begin{aligned}
h_1\vee h_2\colon \setOfFunctionalities{}&\to \mathsf{A}\setOfResources{}\\
\F{f}&\mapsto \Min_{\preceq_{\setOfResources{}}}(h_1(\F{f})\cup h_2(\F{f})).
\end{aligned}
\end{equation*}
\end{definition}

\paragraph*{Algorithm sketch}
Given the aforementioned discoveries, the algorithmic procedure to solve co-design problems is the following.
1) Take an arbitrary \gls{abk:cdpi} (inrerconnection of \glspl{abk:mdpi}). 2) Flatten it to a graph. 3) Re-write the graph in form of a series of series-parallel and feedback graphs. 4) Write the graph as a tree of composition operations. 5) Run Kleene's iteration recursively on the graph.

\subsubsection{Complexity of the solution}
The results for complexity are described in \cite{Censi2015}.
Consider a \gls{abk:cdpi}.
The space of the solution is bounded by the width of $\setOfResources{}$. 
At each iteration, the number of evaluations of each component is linear in the number of options.
The number of execution steps depends on the height of the poset of antichains of $\setOfResources{}$.}

\subsection{Proofs}

\begin{proof}[Proof of Lemma~\ref{lem:funposet}]
Consider partial orders~$A,B,C$ and maps~$f,g,h\colon A\to B$. Clearly~$f\preceq_{B^A}f$. Furthermore, if~$f\preceq_{B^A}g$ and~$g\preceq_{C^B}h$ (i.e.,~$f(a)\preceq_B g(a)$ and~$g(a)\preceq h(a)$,~$\forall a\in A$), then~$f(a)\preceq_B h(a) \ \forall a\in A$, implying~$f\preceq_{B^A}h$. 
Finally, if~$f\preceq_{B^A}g$ and~$g\preceq_{B^A}f$, one has~$f=g$.
\end{proof}

\begin{proof}[Proof of Lemma~\ref{lem:graphposet}]
Consider~$\multigraph_1,\multigraph_2,\multigraph_3\in \mathbf{G}$ with~$\multigraph_1=\tup{\setOfVertices_1,\setOfArcs_1,c_1}$,~$\multigraph_2=\tup{\setOfVertices_2,\setOfArcs_2,c_2}$,~$\multigraph_3=\tup{\setOfVertices_3,\setOfArcs_3,c_3}$, and~$c_i\colon \setOfArcs_i\to C$.
Clearly~$\multigraph_1\preceq_\setofGraphs \multigraph_1$, since~$\setOfVertices_1\subseteq \setOfVertices_1$,~$\setOfArcs_1\subseteq \setOfArcs_1$, and~$c_1\succeq_{C^{\setOfArcs_1}}c_1$. Furthermore, given~$\multigraph_1\preceq_\setofGraphs \multigraph_2$ and~$\multigraph_2\preceq_\setofGraphs \multigraph_3$ (i.e.,~$\setOfVertices_1\subseteq\setOfVertices_2\subseteq \setOfVertices_3$,~$\setOfArcs_1\subseteq\setOfArcs_2\subseteq \setOfArcs_3$,~$c_1\succeq_{C^{\setOfArcs_1}}c_2|_{\setOfArcs_1}$, and~$c_2\succeq_{C^{\setOfArcs_2}}c_3|_{\setOfArcs_2}$), one has~~$\setOfVertices_1\subseteq \setOfVertices_3$,$\setOfArcs_1\subseteq\setOfArcs_3$, and~$c_1\succeq_{C^{\setOfArcs_1}} c_3|_{\setOfArcs_1}$, implying~$\multigraph_1\preceq_\setofGraphs \multigraph_3$. Finally, it is easy to see that~$\multigraph_1\preceq_\setofGraphs \multigraph_2$ and~$\multigraph_2\preceq_\setofGraphs\multigraph_1$ implies~$\multigraph_1=\multigraph_2$.
\end{proof}

\begin{proof}[Proof of Lemma~\ref{lem:poset_travel_isposet}]
Consider~$Q_1,Q_2,Q_3\in \mathcal{Q}$. Clearly~$Q_1\preceq_\mathcal{Q}Q_1$.
Let $Q_1\preceq_\mathcal{Q}Q_2$ and $Q_2\preceq_\mathcal{Q}Q_3$, and let $\tup{o^1,d^1,\alpha^1}\in Q_1$. Since $Q_1\preceq_\mathcal{Q}Q_2$, there is $\tup{o^2,d^2,\alpha^2}\in Q_2$ such that $o^1=o^2$, $d^1=d^2$, and $\alpha^2\geq\alpha^1$. Since $Q_2\preceq_\mathcal{Q}Q_3$, there is $\tup{o^3,d^3,\alpha^3}\in Q_3$ such that $o^2=o^3$, $d^2=d^3$, and $\alpha^3\geq\alpha^2$. So, $o^1=o^3$, $d^1=d^3$, $\alpha^3\geq\alpha^1$, proving that $Q_1\preceq_\mathcal{Q} Q_3$. Finally, $Q_1\preceq_\mathcal{Q} Q_2$ and $Q_2\preceq_\mathcal{Q} Q_1$ implies $Q_1=Q_2$ (given that origin-destination pairs are not repeated).
\end{proof}

\begin{proof}[Proof of Lemma~\ref{lem:red_road_monotone}]
We need to prove that for~$v_1,v_2\in \posreals$ one has:~$v_1\leq v_2\Rightarrow \mathrm{red}_\mathrm{R,V}(v_1)\preceq_\setofGraphs \mathrm{red}_\mathrm{R,V}(v_2)$. Following the definition,~$\mathrm{red}_\mathrm{R,V}(v_1)$ and~$\mathrm{red}_\mathrm{R,V}(v_2)$ will share the same set of vertices (satisfying the vertex condition). Furthermore,~$v_1\leq v_2$ implies that the arcs~$\setOfArcs_1$ of~$\mathrm{red}_\mathrm{R,V}(v_1)$ will be a subset of the set of arcs~$\setOfArcs_2$ of~$\mathrm{red}_\mathrm{R,V}(v_2)$. Finally, the edge colors remain unchanged, except for speed-related one. Let~$c_1,c_2$ the colors associated to~$\mathrm{red}_\mathrm{R,V}(v_1)$ and~$\mathrm{red}_\mathrm{R,V}(v_2)$, respectively. Clearly~$\min\{v_1,x\}\leq \min\{v_2,x\}$ for any~$x\in \posreals$. This, together with \cref{eq:red_condition_road_veh}, gives~$c_1\succeq_{C^{\setOfArcs_1}} c_2$, proving monotonicity.
\end{proof}

\begin{proof}[Proof of Lemma \ref{lem:av_dp_welldef}]
$\R{\costFixVeh},\R{\costOpVeh}$ are monotone functions of the \gls{abk:av}'s achievable speed. Leveraging Lemma~\ref{lem:red_road_monotone}, we know that the serviced network is a monotone function of the speed.
\end{proof}

\begin{proof}[Proof of Lemma~\ref{lem:red_veh_micro_monotone}]
We need to prove that given~$v_1,v_2\in \posreals$, one has:~$v_1\leq v_2\Rightarrow \mathrm{red}_\mathrm{R,M}(v_1)\preceq_\setofGraphs \mathrm{red}_\mathrm{R,M}(v_2)$. First, notice that sets of vertices and arcs are preserved by~$\mathrm{red}_\mathrm{R,M}$. 
Second, the argument for the edge attributes is analogous to the one in the proof of Lemma~\ref{lem:red_road_monotone}. The two facts together prove monotonicity.
\end{proof}

\begin{proof}[Proof of Lemma~\ref{lem:dp_mm_welldef}]
$\R{\costFixSco},\R{\costOpSco}$ are monotone functions of the \gls{abk:mmveh}'s achievable speed. Leveraging Lemma~\ref{lem:red_veh_micro_monotone}, we know that the serviced network is a monotone function of the speed.
\end{proof}

\begin{proof}[Proof of Lemma~\ref{lem:red_pub_monotone}]
We need to prove that given~$n_1,n_2\in \posreals$, one has:~$n_1\leq n_2\Rightarrow \mathrm{red}_\mathrm{P}(n_1)\preceq_\setofGraphs \mathrm{red}_\mathrm{P}(n_2)$. Again, we notice that the set of vertices and arcs are preserved by~$\mathrm{red}_\mathrm{P}$. 
Furthermore,~$n_1\leq n_2$ implies that~$\timePedestrianSubway+ \frac{\numberFleetTrainBaseline}{2n_1\freqTrainBaseline}\geq \timePedestrianSubway+ \frac{\numberFleetTrainBaseline}{2n_2\freqTrainBaseline}$, proving monotonicity.
\end{proof}

\begin{proof}[Proof of Lemma~\ref{lem:dp_pub_welldef}]
First, notice that~$\R{\costSub}$ and~$\R{\emissionsSub}$ are monotone functions of~$\numberFleetTrain$. Furthermore, leveraging Lemma~\ref{lem:red_pub_monotone}, we know that the serviced network relates monotonically to~$\numberFleetTrain$.
\end{proof}

\begin{proof}[Proof of Lemma~\ref{lem:iamod_1_dp_welldef}]
Let $\R{r}\preceq_{\R{\setofGraphs}^2\times \R{\bar{\mathbb{N}}}\times \R{\rplusbar}^3} \R{r'}$. Since all feasible solutions of \eqref{eq:TIamodopt_1} with $\R{r}$ remain feasible with $\R{r'}$, $\bar{d}_\mathrm{IAMOD}(\F{q}^*,\R{r})\subseteq \bar{d}_\mathrm{IAMOD}(\F{q}^*,\R{r'})$ for all $\F{q}\in\F{\mathcal{Q}}$.
Similarly, let~$\F{q'}\preceq_{\F{\mathcal{Q}}} \F{q}$.  Since all feasible solutions of \eqref{eq:TIamodopt_1} remain feasible (possibly by replacing demand with empty vehicles and by artificially adding loops to the graph), $\bar{d}_\mathrm{IAMOD}(\F{q'},\R{r})\subseteq \bar{d}_\mathrm{IAMOD}(\F{q},\R{r})$ for all $\R{r}\in\R{\setofGraphs}^2\times \R{\bar{\mathbb{N}}}\times \R{\rplusbar}^3$.
This proves monotonicity.
\end{proof}

\begin{proof}[Proof of Lemmas~\ref{lem:iamod_2_dp_welldef} and~\ref{lem:iamod_3_dp_welldef}]
The proofs parallel the proof of Lemma~\ref{lem:iamod_1_dp_welldef}.
\end{proof}

\begin{proof}[Proof of Lemmas~\ref{lem:mob_1_welldef},~\ref{lem:mob_2_welldef}, and~\ref{lem:mob_3_welldef}]
The \glspl{abk:mdpi} are monotone, since they consist of the valid composition of monotone \glspl{abk:mdpi}~\cite{Censi2015}.
\end{proof}
\end{appendix}
\bibliographystyle{IEEEtran}
\bibliography{paper.bib}

\begin{IEEEbiography}[{\includegraphics[width=1in,height=1.25in,clip,keepaspectratio]{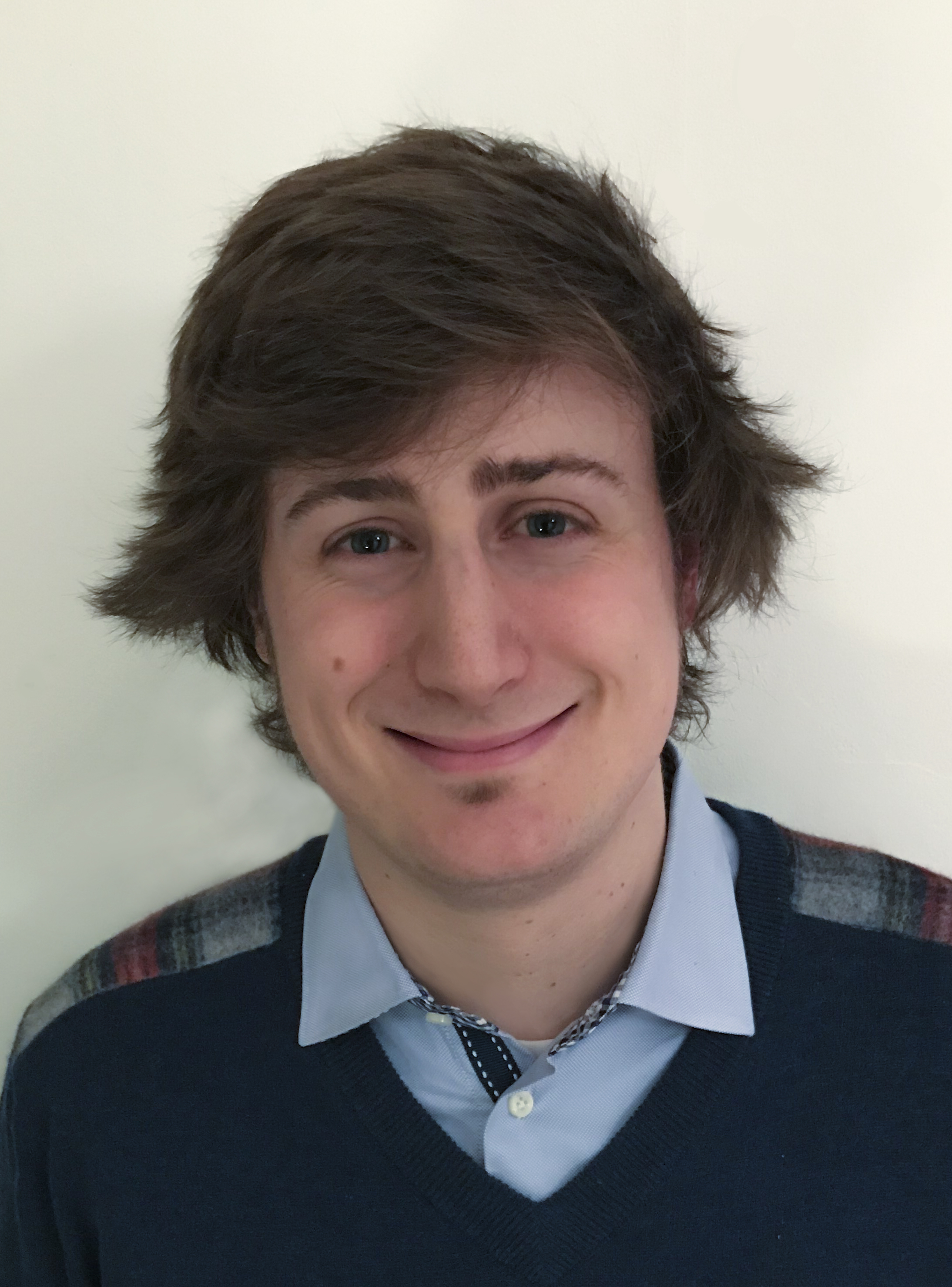}}]{Gioele Zardini (gzardini@ethz.ch)} 
is a Ph.D. candidate at the Institute for Dynamic Systems and Control at ETH Zurich, under the supervision of Prof. Emilio Frazzoli.
He received the B.Sc. and the M.Sc. degrees in mechanical engineering, with focus in Robotics, Systems, and Control from ETH Zurich in 2017 and 2019, respectively.
He worked at nuTonomy (then Aptiv AM, now Motional) and was a visiting researcher at Stanford University and Massachusetts Institute of Technology.
His current research interests include the co-design of complex systems (all the weay from future mobility to embodied intelligence), compositionality in engineering, planning and control, and game theory.
He is the recipient of the Best Paper Award (1st Place) at the 2021 IEEE International Conference on Intelligent Transportation Systems.
\end{IEEEbiography}

\begin{IEEEbiography}[{\includegraphics[width=1in,height=1.25in,clip,keepaspectratio]{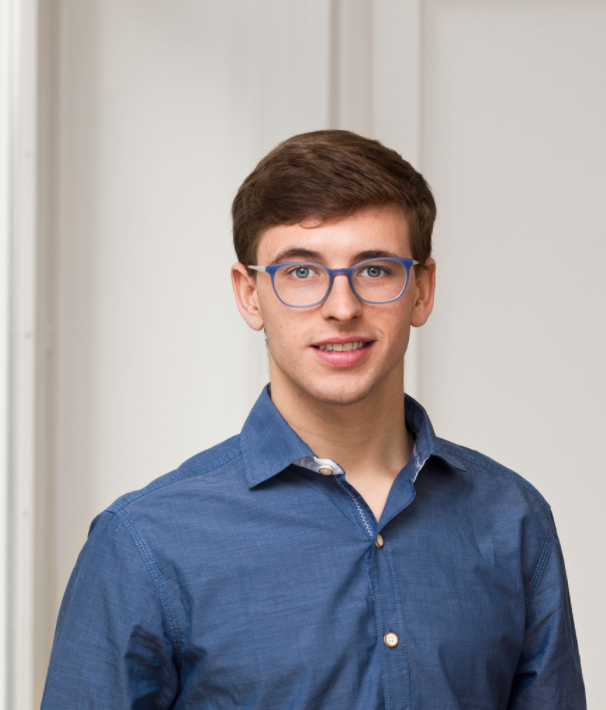}}]{Nicolas Lanzetti (lnicolas@ethz.ch)} 
is a Ph.D. candidate at the Automatic Control Laboratory at ETH Zurich, under the supervision of Prof. Florian D\"orfler.
He received the B.Sc. and the M.Sc. degrees in mechanical engineering, with focus in Robotics, Systems, and Control from ETH Zurich in 2016 and 2019, respectively.
He was a visiting researcher at Massachusetts Institute of Technology and Stanford University.
His current research interests include optimal transport and gradient flows in the Wasserstein space, with applications in robust optimization and game theory.
He is the recipient of the Willi Studer Prize, the ETH Medal and the SVOR/ASRO award for his Master's thesis, and the Best Paper Award (1st Place) at the 2021 IEEE International Conference on Intelligent Transportation Systems.
\end{IEEEbiography}

\begin{IEEEbiography}[{\includegraphics[width=1in,height=1.25in,clip,keepaspectratio]{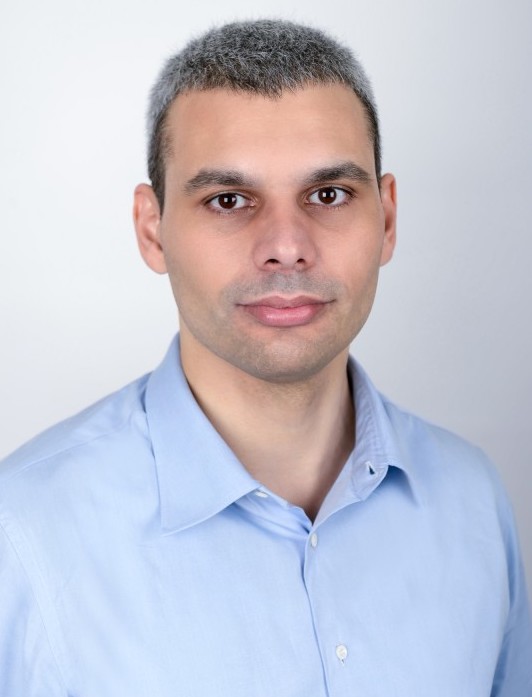}}]{Andrea Censi (acensi@ethz.ch)} 
is the deputy director of the Dynamic Systems and Control chair at ETH Zurich, director of the Duckietown Foundation, and  founder of Zupermind.
He obtained a M.Eng. degree in Control and Robotics from the University of Rome, ``Sapienza'', and a Ph.D. from California Institute of Technology. He has been a research scientist at the Massachusetts Institute of Technology, and the Director of Research at Aptiv Autonomous Mobility (now Motional).
He has been the recipient of NSF and AFRL awards.
\end{IEEEbiography}

\begin{IEEEbiography}[{\includegraphics[width=1in,height=1.25in,clip,keepaspectratio]{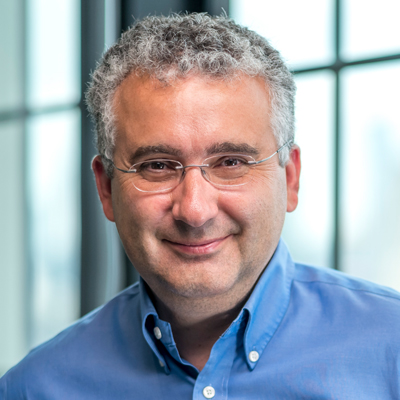}}]{Emilio Frazzoli (efrazzoli@ethz.ch)} is a Professor of Dynamic Systems and Control at ETH Zurich. 
Until March 2021, he was Chief Scientist of Motional, the latest embodiment of nuTonomy, the startup he founded with Karl Iagnemma in 2013.
He received the Laurea degree in aerospace engineering from the University of Rome, ``Sapienza'', in 1994, and the Ph.D. degree in Aeronautics and Astronautics from the Massachusetts Institute of Technology in 2001.
Before joining ETH Zurich in 2016, he held faculty positions at the University of Illinois, Urbana Champaign, the University of California, Los Angeles, and at the Massachusetts Institute of Technology.
His current research interests focus primarily on autonomous vehicles, mobile robotics, and transportation systems.
He led the research groups that first demonstrated an autonomous mobility service to the public, and performed the first analysis of the social and economic impact of such a service, based on real transportation data.

He was the recipient of a NSF CAREER award in 2002, the IEEE George S. Axelby award in 2015, the IEEE Kiyo Tomiyasu award in 2017, and has been named an IEEE Fellow in 2019.
\end{IEEEbiography}

\begin{IEEEbiography}[{\includegraphics[width=1in,height=1.25in,clip,keepaspectratio]{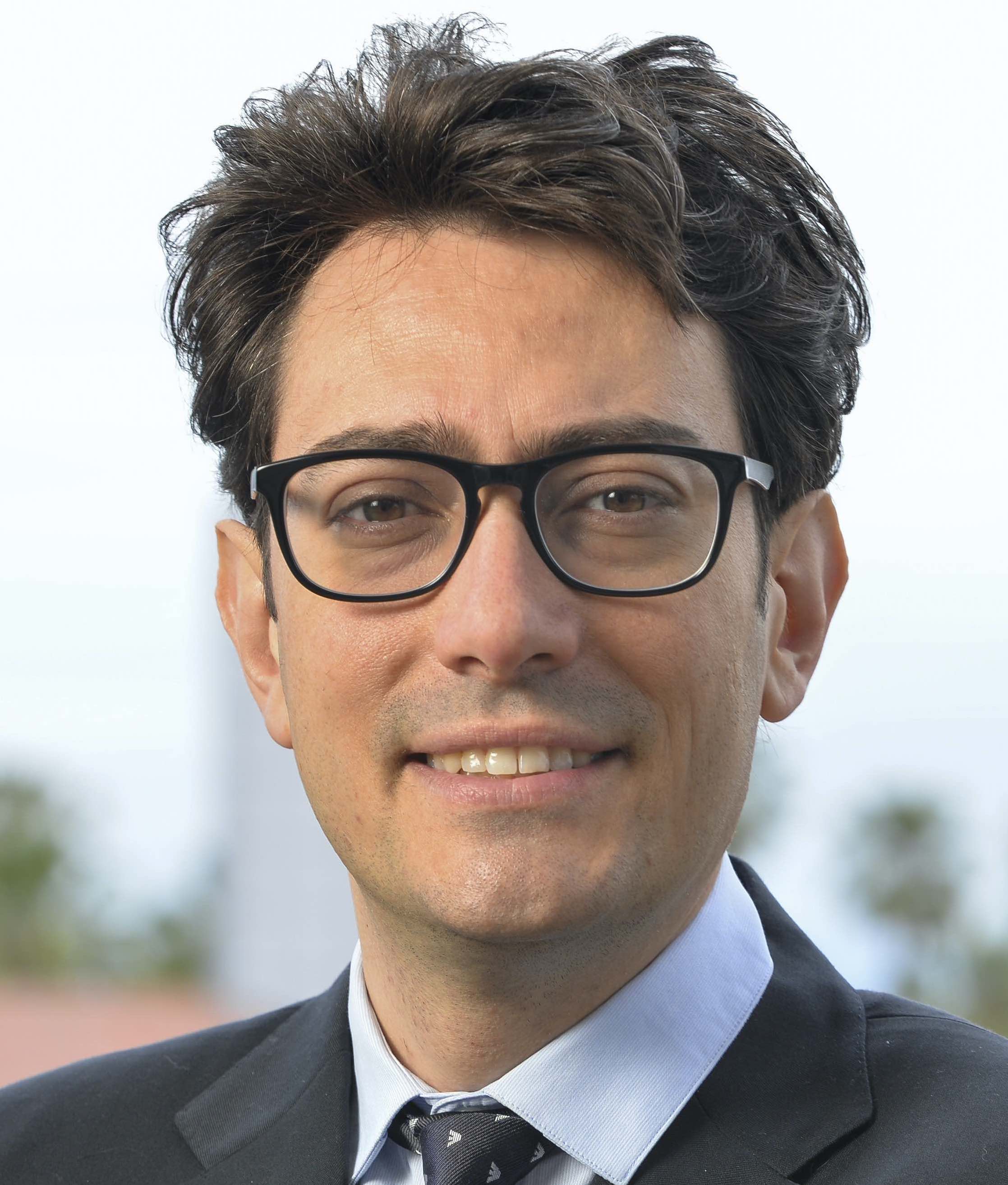}}]{Marco Pavone (pavone@stanford.edu)} is an Associate Professor of Aeronautics and Astronautics at Stanford University, where he is the Director of the Autonomous Systems Laboratory and Co-Director of the Center for Automotive Research at Stanford. He is currently on a partial leave of absence at NVIDIA serving as Director of Autonomous Vehicle Research. He received a Ph.D. degree in Aeronautics and Astronautics from the Massachusetts Institute of Technology in 2010. His main research interests are in the development of methodologies for the analysis, design, and control of autonomous systems, with an emphasis on self-driving cars, autonomous aerospace vehicles, and future mobility systems. 
He is a recipient of a number of awards, including a Presidential Early Career Award for Scientists and Engineers from President Barack Obama, an Office of Naval Research Young Investigator Award, a National Science Foundation Early Career (CAREER) Award, a NASA Early Career Faculty Award, and an Early-Career Spotlight Award from the Robotics Science and Systems Foundation. He was identified by the American Society for Engineering Education (ASEE) as one of America's 20 most highly promising investigators under the age of 40.
He is currently serving as an Associate Editor for the IEEE Control Systems Magazine.
\end{IEEEbiography}

\vfill
\end{document}